\documentclass[10pt,twocolumn,superscriptaddress,letterpaper,prl]{revtex4}

\usepackage[usenames,dvipsnames]{xcolor}

\usepackage{lscape}
\usepackage{booktabs,dcolumn}
\usepackage{graphicx,amsmath,color,amssymb}
\usepackage[latin1]{inputenc}
\usepackage[sort&compress]{natbib}
\usepackage[version-1-compatibility]{siunitx}
\definecolor{darkblue}{rgb}{0,0,.4}
\definecolor{darkred}{rgb}{0.6,0,0}

\newcommand{\abs}[1]{\left|#1\right|}

\newcommand{\ket}[1]{\left| #1 \right>}

\renewcommand\eqref[1]{Eq.~(\ref{#1})}

\newcommand{\rr}{\mathbf{r}}
\newcommand{\RR}{\mathbf{R}}
\newcommand{\me}[3]{\left< #1 \left| #2 \right| #3 \right>}
\newcommand{\ovI}[2]{\left< #1 | #2 \right>}

\usepackage[bookmarksopen=true, % Lesezeichen am linken Rand
citecolor=blue,
colorlinks=true, % farbige Links, kein Rahmen
linkcolor=blue]{hyperref} % Verlinkungen

\begin{document}
%\lipsum[1]

\title{Single-photon superradiance from a quantum dot}

\author{Petru Tighineanu}
\email{petrut@nbi.ku.dk}
\affiliation{Niels Bohr Institute,\ University of Copenhagen,\ Blegdamsvej 17,\ DK-2100 Copenhagen,\ Denmark}
\author{Rapha\"el S. Daveau}
\affiliation{Niels Bohr Institute,\ University of Copenhagen,\ Blegdamsvej 17,\ DK-2100 Copenhagen,\ Denmark}
\author{Tau B. Lehmann}
\affiliation{Niels Bohr Institute,\ University of Copenhagen,\ Blegdamsvej 17,\ DK-2100 Copenhagen,\ Denmark}
\author{Harvey E.~Beere}
\affiliation{Cavendish Laboratory,\ University of Cambridge,\ J.\ J.\ Thomson Avenue,\ CB3 0HE Cambridge,\ UK}
\author{David A.~Ritchie}
\affiliation{Cavendish Laboratory,\ University of Cambridge,\ J.\ J.\ Thomson Avenue,\ CB3 0HE Cambridge,\ UK}
\author{Peter Lodahl}
\affiliation{Niels Bohr Institute,\ University of Copenhagen,\ Blegdamsvej 17,\ DK-2100 Copenhagen,\ Denmark}
\author{S{\o}ren Stobbe}
\email{stobbe@nbi.ku.dk}
\affiliation{Niels Bohr Institute,\ University of Copenhagen,\ Blegdamsvej 17,\ DK-2100 Copenhagen,\ Denmark}

\begin{abstract}
We report on the observation of single-photon superradiance from an exciton in a semiconductor quantum dot. The confinement by the quantum dot is strong enough for it to mimic a two-level atom, yet sufficiently weak to ensure superradiance. The electrostatic interaction between the electron and the hole comprising the exciton gives rise to an anharmonic spectrum, which we exploit to prepare the superradiant quantum state deterministically with a laser pulse. We observe a five-fold enhancement of the oscillator strength compared to conventional quantum dots. The enhancement is limited by the base temperature of our cryostat and may lead to oscillator strengths above 1000 from a single quantum emitter at optical frequencies.
\end{abstract}

\maketitle

\begin{figure}[t!]
\includegraphics[width=\columnwidth]{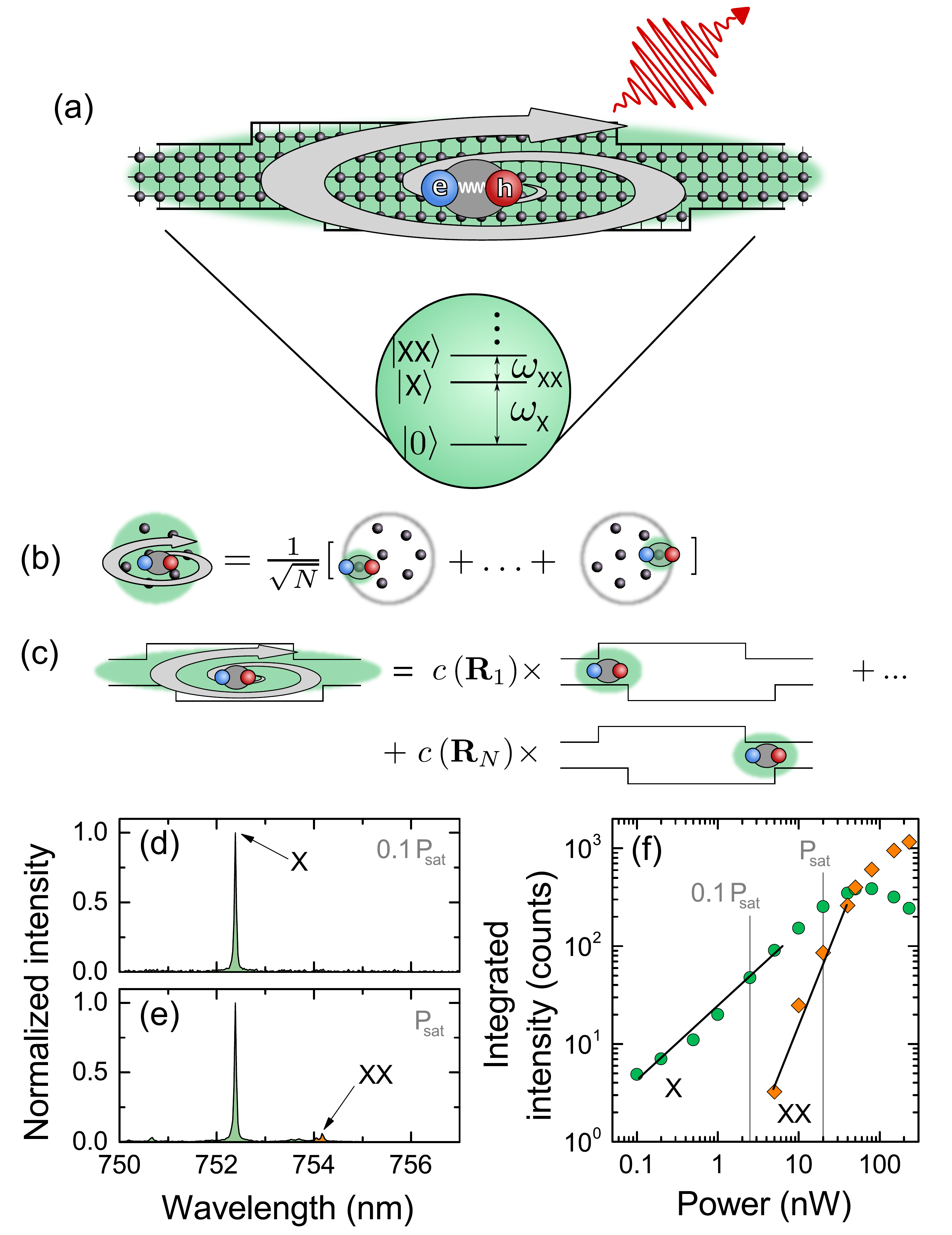}
\caption{\label{Figure1} Superradiant excitons in quantum dots. (a) A quantum dot defined by intentional monolayer fluctuations weakly confine electrons (e) and holes (h), which are mutually bound by electrostatic attraction. Notably, the spectrum is anharmonic due to interactions, i.e., the energy $\hbar \omega_\text{XX}$ of a biexciton is less than the energy $\hbar \omega_\text{X}$ of a single exciton. The swirling arrow indicates superradiantly enhanced light-matter coupling. (b) SPS is defined in an ensemble of non-interacting emitters as a symmetric superposition of different excitations. (c) The excitonic enhancement of light-matter interaction may be regarded as a generalization of SPS: the exciton is a symmetric superposition of excitations. (d) Measured photoluminescence spectrum at 10\% of the exciton saturation power $P_\text{sat} = \SI{20}{\nano\watt}$. Only the exciton is observed. (e) At the saturation power, the biexciton becomes discernible. (f) The excitons and biexcitons are distinguished by their power-law dependence on excitation power, $P$: the fits yield $P^{0.86}$ and $P^{2.01}$, respectively.
}
\end{figure}

Enhancing and tailoring light-matter interaction is at the heart of modern quantum physics, partly because it enables studying hitherto unexplored realms of physics and partly to meet the steep requirements for quantum-information science. Photonic nanostructures efficiently tailor the density of optical states and have proven very useful to this end. For example, cavities can reach strong coupling to emitters~\cite{Yoshie2004Nature,Reithmaier2004Nature} or mechanical objects~\cite{Groblacher2009Nature}, and photonic waveguides enable efficient photonic switches~\cite{Tiecke2014Nature} and single-photon sources~\cite{Arcari2014PRL}. Another approach to enhancing light-matter interaction concerns tailoring the capability of the emitter to be polarized, i.e., the oscillator strength. This can be achieved with collective effects such as superradiance~\cite{dicke54}, which has been studied in ensembles of atoms~\cite{Haroche2013RMP}, ions~\cite{devoe96}, Bose-Einstein condensates~\cite{Baumann2010Nature}, and superconducting circuits~\cite{Mlynek2014NCOMM}. Collective enhancement can occur at the single-photon level if a single quantum of energy is distributed coherently in an ensemble~\cite{dicke54}. This single-photon superradiance (SPS) has been studied so far in ensembles of non-interacting emitters such as nuclei~\cite{rohlsberger10}, and is central to schemes for robust quantum communication~\cite{DLCZ} and quantum memories~\cite{Hammerer2010RMP}. A drawback of non-interacting systems is their harmonic energy structure, which prohibits deterministic preparation of a particular collective state. Here we show that the fundamental optical excitation of a weakly confining quantum dot is a generalization of SPS. We prepare the collective quantum state deterministically with a laser pulse and demonstrate its superradiant character. Our findings underline the extraordinary potential of weakly confining quantum dots for achieving unprecedented light-matter coupling strengths at optical frequencies, which would improve the radiative efficiency, quantum efficiency, quantum nonlinearities, and coherence of single-photon sources in nanophotonic quantum devices~\cite{Lodahl2014RMP}.

% Currently employed quantum emitters operating at optical frequencies suffer from small oscillator strengths ranging between 1 and 15

We study quantum dots formed by intentional monolayer fluctuations of a quantum well, which were pioneered by Gammon et al.~\cite{gammon96}, cf.~Fig.~\ref{Figure1}(a). The subwavelength size of the quantum dot is key to achieving a large collective enhancement; in larger ensembles, such as atomic clouds, the enhancement is reduced by destructive interference~\cite{dicke54,scully09super}. The fundamental optical excitation of the quantum dot is an electron-hole pair bound by electrostatic attraction and quantum confinement, i.e., an exciton. We demonstrate that the exciton recombines radiatively with a quantum efficiency of \SI{99(2)}{\percent}, which is the highest reported on quantum dots so far~\cite{Johansen2008PRB,Stobbe2010PRB,tighineanu13}. The resulting single photons inherit the superradiant character in the form of an enhanced emission rate compared to conventional strongly confining quantum dots. We employ a recently developed method exploiting the fine structure of the exciton~\cite{tighineanu13} and measure an oscillator strength of up to $96\pm2$. The corresponding superradiant enhancement of about 5 is limited by the base temperature of our cryostat (\SI{7}{\kelvin}) and could potentially be orders of magnitude larger at temperatures below \SI{1}{\kelvin}.

The hallmark of SPS is the symmetric collective quantum state~\cite{dicke54},
\begin{equation}
\ket{\Psi} = \frac{1}{\sqrt{N}} \sum_j\ket{g_1g_2...e_j...g_N},
\label{Equation1}
\end{equation}
where $N$ is the number of emitters, the $j$-th emitter is in the excited state $\ket{e}$ and all others are in the ground state $\ket{g}$. The remarkable property of $\ket{\Psi}$ is that it interacts with light $N$ times stronger than a single emitter. This state describes a non-interacting ensemble, where the excitation is localized in a single emitter at a time as depicted in Fig.~\ref{Figure1}(b). In a system of interacting particles, such as a semiconductor quantum dot, the wavefunctions of the underlying atoms overlap, leading to delocalized excitations. This destroys the collective enhancement of light-matter interaction and causes conventional quantum dots to exhibit small oscillator strengths of about 10, despite that they embody tens of thousands of atoms.

The spatial extent of delocalized excitations is a fundamental property of semiconductors and is determined by the size of an exciton. Enhancement of light-matter interaction can therefore be achieved only in quantum dots that are larger than the exciton radius. This regime is known as weak confinement and the enhancement of the light-matter coupling in weakly confining quantum dots was first predicted by Hanamura~\cite{hanamura88} and dates back to early theoretical studies of impurities in semiconductors~\cite{rashba62}. Here we show that this effect is equivalent to SPS, and the exciton state can be written as, see Supplementary information~\cite{SIprl},
\begin{equation}
\Psi_\mathrm{X}(\RR,\rr) = \sum_j c(\RR_j)\phi_\mathrm{X}(\RR-\RR_j,\rr),
\label{Equation2}
\end{equation}
where $\rr$ ($\RR$) is the relative (center-of-mass) electron-hole coordinate, and the index $j$ runs over the unit cells constituting the quantum dot. The function $\phi_\mathrm{X}$ describes an exciton with the size and oscillator strength of a conventional quantum dot, while $c$ is responsible for the collective enhancement as illustrated in Fig.~\ref{Figure1}(c). The light-matter coupling is proportional to the number of atoms comprising the weakly confining quantum dot. This analysis shows that this effect is a generalization of SPS, cf.\ Figs.~\ref{Figure1}(b,c), and the two effects are equivalent if $c$ is constant throughout the quantum dot. The constant phase of $c$ found for ground-state excitons with $s$-like symmetry ensures constructive interference among the excitations defined by $\phi_\mathrm{X}$.

An exciton governed by Eq.\ (\ref{Equation2}) has been long sought in solid-state quantum optics~\cite{Andreani1999PRB} because it can lead to large oscillator strengths. Realizing weakly confining quantum dots has been a challenge so far because it requires precise control over growth parameters to obtain a homogeneous potential profile over extended length scales. Previous studies on large quantum dots~\cite{Stobbe2010PRB,tighineanu13} revealed small oscillator strengths, which is believed to be caused by inhomogeneous potential profiles within the quantum dots. Previous works~\cite{guest02,Hours2005PRB} found fast recombination rates in gallium-arsenide quantum dots but without rigorous information about the impact of non-radiative processes, see Supplementary information~\cite{SIprl} for further discussions. The measured spectra shown in Fig.~\ref{Figure1}(d) and (e) were obtained by exciting in the quasi-continuum energy band of the quantum well as discussed below. An exciton and a biexciton are identified as shown in Fig.~\ref{Figure1}(f). These quasi-particles radiate at different frequencies, cf.~Fig.~\ref{Figure1}(e), which reflects the spectral anharmonicity of the quantum dot.

\begin{figure}
\includegraphics[width=\columnwidth]{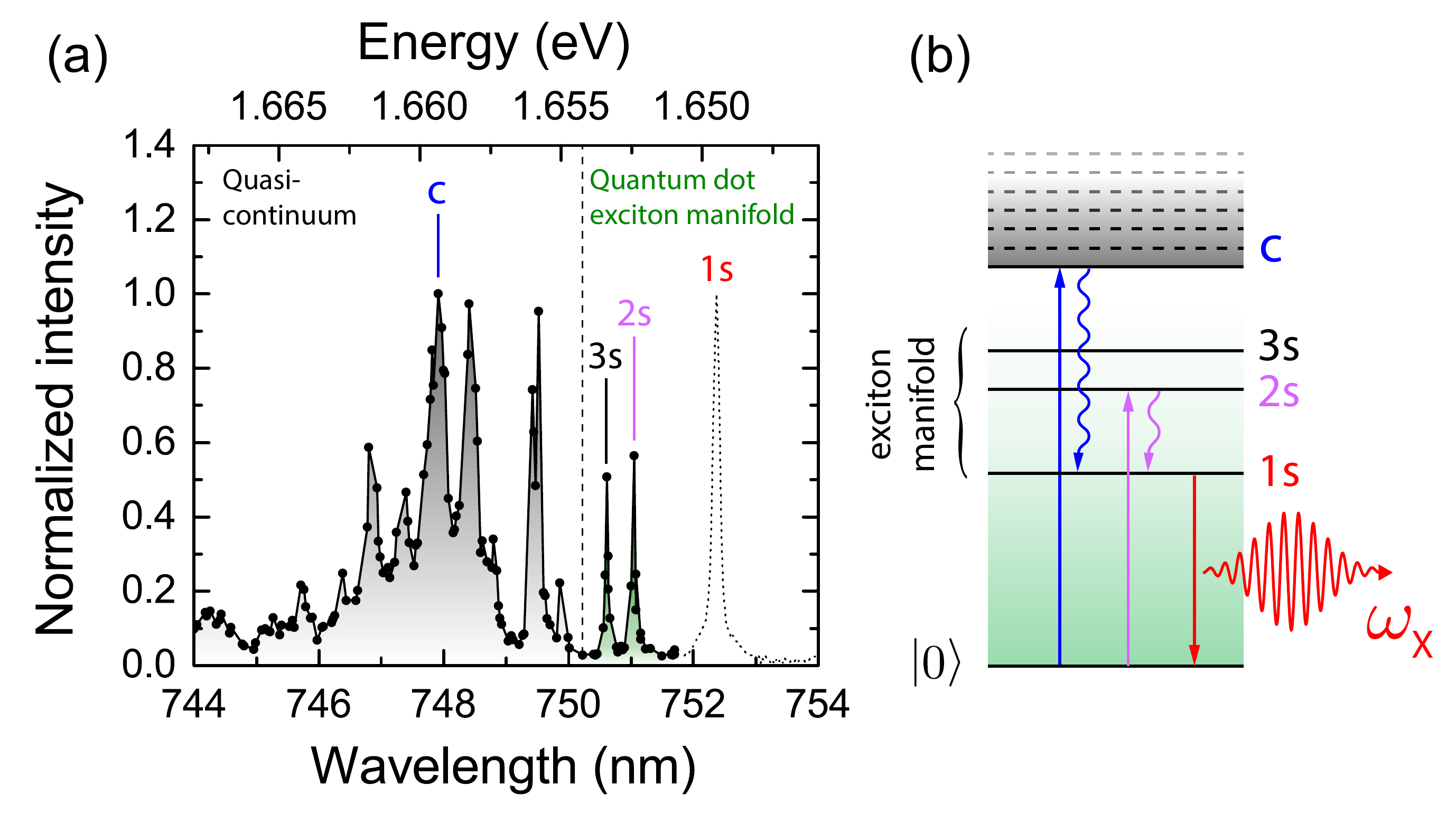}
\caption{\label{Figure2} Deterministic preparation of superradiant excitons. (a) Photoluminescence-excitation spectrum obtained by integrating the emission of the 1s transition while scanning the excitation wavelength. It features a quasi-continuum and resolves the lowest-energy states of the exciton manifold, labeled 1s, 2s, and 3s. (b) Two excitation schemes are used in our study. With C-type excitation, an equal bright- and dark-exciton population is prepared, which is important for extracting the impact of non-radiative processes. Deterministic preparation of the bright exciton is achieved by pumping into the 2s state.}
\end{figure}

\begin{figure*}[t]
\includegraphics[width=\textwidth]{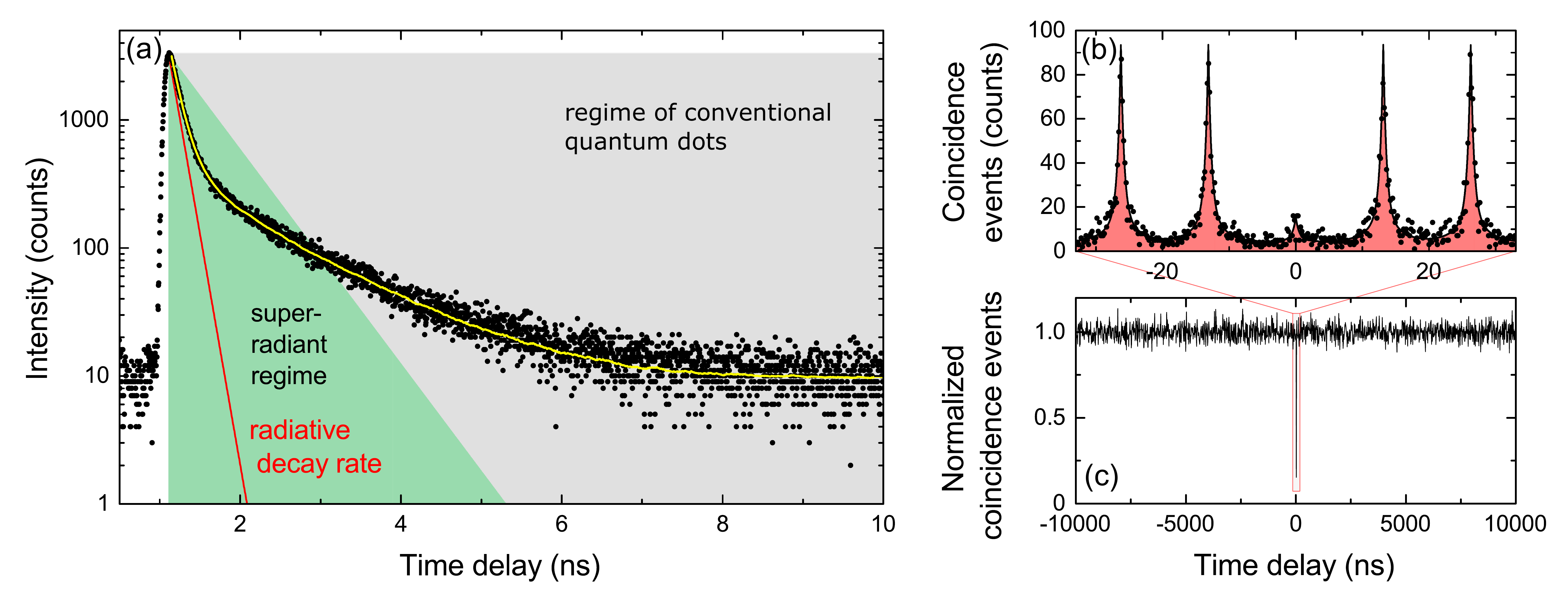}
\caption{\label{Figure3} Experimental demonstration of single-photon superradiance from a quantum dot. (a) Time-resolved decay (black points) of the bright 1s exciton obtained under 2s excitation. We obtain an excellent fit (yellow line) to the theoretical model when convoluting with the instrument-response function of the detector. After separating non-radiative from radiative effects we extract a radiative decay rate of \SI{8.3(1)}{\nano\second^{-1}} (red line), which is deeply in the superradiant regime (green area). (b) HBT measurement of the emitted photons showing $g^{(2)}(0)=0.13$, which proves single-photon emission. (c) Long-time-scale HBT measurement where each coincidence peak has been numerically integrated. No blinking is observed.}
\end{figure*}

To identify proper excitation conditions of the quantum dot, we probe the spectrum of states with photoluminescence-excitation spectroscopy as displayed in Fig.~\ref{Figure2}(a). The spectrum shows a quasi-continuum band of quantum-dot states hybridized with quantum-well states as well as the exciton manifold in which we identify the 1s, 2s, and 3s states of two-dimensional excitonic hydrogen~\cite{Que1992PRB}. Key features of the spectrum are summarized in Fig.~\ref{Figure2}(b). We use two excitation conditions to prepare the 1s exciton: i)~Pumping in the quasi-continuum band of states (C-type excitation) allows extracting the impact of non-radiative processes governing the 1s exciton decay. Since the quantum dot traps carriers with random spin, equal populations of spin-bright and spin-dark 1s excitons are prepared. While only bright excitons emit light, the dark excitons influence the decay dynamics and play a key role in revealing non-radiative effects (see Supplementary information~\cite{SIprl}). ii)~Deterministic preparation of spin-bright superradiant 1s excitons is achieved by pumping into the 2s exciton state, cf. Fig.~\ref{Figure2}, with a pulsed laser. This is feasible since the decay cascade from 2s to 1s is spin-conserving~\cite{Poem2010NPHYS} and spin-dark states are not populated. Deterministic excitation occurs when applying sufficient optical power (\SI{300}{\nano\watt}) to saturate the emission from the 1s state.

The figure of merit for collective enhancement of light-matter interaction is the oscillator strength, $f$, which gauges the strength of the interaction with light. The oscillator strength is determined by the radiative spontaneous-emission rate of an emitter placed in a homogenous photonic environment. In an experiment, however, the oscillator strength is masked by non-radiative effects and the non-homogeneity of the photonic environment, whose contributions are fully addressed in our study, see Supplementary information~\cite{SIprl} for further details. Central to our analysis is a recently developed method exploiting the fine structure of excitons to rigorously separate radiative from non-radiative effects~\cite{Wang2011PRL}. Figure \ref{Figure3}\textbf{a} shows raw data of the time-resolved decay of the deterministically prepared 1s exciton. We measure an excellent near-unity radiative efficiency of $\eta=(99\pm2)\%$, which is the highest ever measured on quantum dots so far. The extracted oscillator strength of $f=72.0\pm0.8$ is enhanced far beyond the upper limit of $f=17.4$ for conventional quantum dots at this wavelength. By combining structural information about the sample with the measured oscillator strength we can faithfully reconstruct the exciton wave function and find a diameter of $\SI{24}{\nano\meter}$, which is smaller than the wavelength of light yet sufficiently large to embody $\sim 90,000$ atoms in a collective quantum state sharing a single quantum of energy.

Figure \ref{Figure3}(b) shows the second-order correlation function obtained in a Hanbury-Brown-Twiss (HBT) experiment from which we find a normalized zero-time correlation function of $g^{(2)}(0)=0.13$. Antibunched emission directly pinpoints to the presence of a single quantum of matter inside the quantum dot, which is an essential property of SPS, cf.~\eqref{Equation1}. In conjunction with the measured enhanced oscillator strength for a spatially confined exciton, this is the unequivocal demonstration of SPS in a quantum dot. 
%We note, however, that antibunched emission is inherent to any anharmonic quantum emitter, including conventional quantum dots, and cannot be used alone as a measure of SPS. 
Solid-state quantum light sources often suffer from charge traps that switch the emitter into an optically dark state, also known as blinking processes, and reduce the preparation efficiency of bright states. This can be quantified from HBT correlations acquired over long time scales as shown in Fig.~\ref{Figure3}(c). No bunching effects are observed, which shows that this single-photon source is free from blinking on a time scale of at least \SI{10}{\micro\second}.

%The anharmonicity of the quantum-dot spectrum confers a pronounced robustness to the superradiant nature of photons emitted by the ground-state exciton. While multi-photon emission destroys SPS in a harmonic emitter leading to a finite $g^{(2)}(0)$, it does not contaminate the single-photon purity of quantum dots because the emitted photons have different color and do not interfere. The only experimentally relevant parameter that may be detrimental to SPS in a quantum dot is temperature, which may lead to populating excited states of the exciton manifold with reduced oscillator strength.

Single-photon superradiance is a robust phenomenon in quantum dots due to the anharmonic spectrum. The only experimentally relevant parameter that may be detrimental to SPS is temperature. For a thermal de Broglie wavelength larger than the quantum-dot size, the exciton Pauli blockage is broken and leads to multi-photon emission, thereby destroying SPS. This non-trivial effect is beyond the scope of the current work and will be presented elsewhere.

We have measured the oscillator strength of 9 quantum dots and found them all to be superradiant with an average oscillator strength of $f=76\pm11$. Further experimental data is included in the Supplementary information~\cite{SIprl}. Remarkably, we have measured a homogeneous-medium radiative decay rate of up to $\Gamma_\mathrm{rad}=\SI{11.1(2)}{\nano\second^{-1}}$, which is the fastest value ever reported for any single-photon source and corresponds to an oscillator strength of $f=\SI{96(2)}{}$. Such a quantum dot can deliver a radiative flux of single photons equivalent to more than five conventional quantum dots.

The superradiant enhancement of the light-matter coupling in quantum dots is proportional to the number of atoms in the collective state, and can potentially be much larger than reported here. The enhancement factor may realistically reach $100\times$ for quantum-dot diameters of $\sim \SI{100}{\nano\meter}$~\cite{Stobbe2012PRB} corresponding to an oscillator strength of $f\sim 1500$. Such highly superradiant quantum dots may exist in our sample but the temperature at which the experiment is carried out ($T=\SI{7}{\kelvin}$) does not allow resolving such large oscillator strengths. This is because in large quantum dots the confinement energy may become smaller than the thermal energy, which results in populating excited states with reduced oscillator strength. The maximum oscillator strength $f_\textrm{max,th}$ that can be resolved at a temperature $T$ is calculated for a quantum dot in which the energy difference between two eigenstates equals $4k_\mathrm{B}T$ (a detailed analysis is provided in the Supplementary information~\cite{SIprl})
%and calculate it by considering a disk-shaped quantum dot in which the energy difference between two energy states is larger than $4k_\mathrm{B}T$
\begin{equation}
f_\textrm{max,th} = \frac{4\hbar E_P}{M\omega a_0^2}\frac{1}{\xi}\frac{1}{k_\mathrm{B}T},
\label{eq:fmaxth}
\end{equation}
where $E_P$ is the Kane energy, $M$ the exciton mass, $a_0$ the exciton radius, $k_\mathrm{B}T$ the thermal energy, and 1:$\xi$ the in-plane aspect ratio of the quantum dot ($\xi \geq 1$). At $T=\SI{7}{\kelvin}$ we find that oscillator strengths larger than $f_\textrm{max,th}= 170$ cannot be resolved for in-plane symmetric quantum dots, and $f_\textrm{max,th}$ decreases even further for more realistic asymmetric shapes. Oscillator strengths of $\sim 1500$ require temperatures below $\sim \SI{0.8}{\kelvin}$. The light-matter coupling may be further enhanced $>10\times$ beyond the homogeneous-medium value by the Purcell effect~\cite{Lodahl2014RMP}. This could allow studying fascinating non-energy-conserving effects such as the ultrastrong regime of light-matter coupling~\cite{Niemcyk2010NPHYS}. The repetition rates of single-photon sources would approach the terahertz regime yielding radiating powers of hundreds of nanowatts from a single quantum emitter. The single-photon emission would potentially be highly coherent, partly due to an intrinsically weaker coupling to nuclear spin noise~\cite{Kuhlmann13} and phonon dephasing~\cite{Rol2007PRB} for large excitons, partly because the dephasing mechanisms present in solid-state environments would be largely negligible compared to a radiative decay at subpicosecond time scales. Even larger decay rates could become possible in materials with small Bohr radii~\cite{Kazimierczuk2014Nature,Srivastava2014}, see~\eqref{eq:fmaxth}. In particular, fast decays have recently been reported in CdSe nanoplatelets~\cite{naeem15}. Another intriguing aspect of the SPS regime is that the collective Lamb shift is predicted to be finite~\cite{scully09} without the renormalization schemes required in the quantum electrodynamics of conventional emitters.

We thank Inah Yeo and Kristian H. Madsen for experimental assistance, and Anders S. S{\o}rensen, J\"{o}rg H. M\"{u}ller and Philip T. Kristensen for theoretical discussions. We gratefully acknowledge the financial support from the Carlsberg Foundation, the Lundbeck Foundation, and the European Research Council (ERC consolidator grant "ALLQUANTUM").

\nocite{takagahara86}
\nocite{cardona05}
\nocite{Ahn2003PRB}
\nocite{bastard84}
\nocite{greene84}
\nocite{bastard84}
\nocite{johansen10}
\nocite{paulus00}
\nocite{Peter2005PRL}
\nocite{Diniz2011PRA}
\nocite{Madsen2013NJP}
\nocite{Reitzenstein2009PRL}
\nocite{Coldren_and_Corzine}
\nocite{Bayer2002PRB}
\nocite{vurgaftman01}
\nocite{barlow89}
\nocite{gregory05}
\nocite{bard74}
\nocite{miller08}
\nocite{sugawara95}

\bibliographystyle{naturemag}
\bibliography{bibliography}

\newpage

\onecolumngrid

\begin{center} \textbf{SUPPLEMENTARY INFORMATION} \end{center}

\tableofcontents

\newpage

\section{I. Theory of single-photon superradiance from quantum dots}
\label{sec:s1}
Quantum dots are well suited for enhancing the light-matter interaction strength due to their multi-atomic nature. However, quantum dots that are smaller than the exciton Bohr radius exhibit a limited oscillator strength because the electron-hole motion is uncorrelated and dominated by quantum-confinement effects. This is known as the strong-confinement regime. In the opposite limit, weak confinement, the relative electron-hole motion is strongly correlated while their center-of-mass motion is bound to the weakly confining potential of the quantum dot \cite{hanamura88}.

\subsection{The strong-confinement regime}
\begin{figure}[t!]
\centering
\includegraphics[width=\columnwidth]{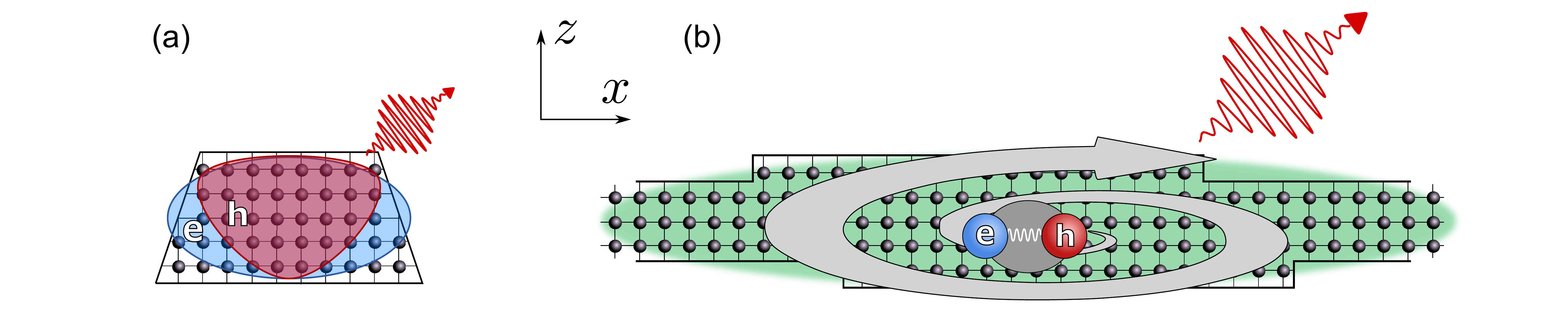}
\caption{ \label{fig:Single_IF_QD} Superradiance with single quantum dots. (a) In small quantum dots, such as self-assembled In(Ga)As quantum dots, the motion of electrons and holes is governed by quantum-confinement effects and is completely uncorrelated, which limits the light-matter interaction strength. (b) In large interface-fluctuation quantum dots, the electron-hole motion is dominated by their mutual attraction in the plane of the quantum dot. The mean electron-hole separation ($\sim$ exciton Bohr radius; dark gray) is smaller than the exciton wave function (green) and leads to a strong superradiant behavior of the ground-state exciton.}
\end{figure}

In the strong-confinement regime, the exciton wave function can be written as a product of the individual wave functions of the electron and the hole, i.e., the electron-hole motion is decoupled as illustrated in Fig.\ \ref{fig:Single_IF_QD}(a). In this limit the oscillator strength is given by \cite{tighineanu13}
\begin{equation}
f=\frac{E_\text{P}}{\hbar\omega}\abs{\ovI{F_h(\rr)}{F_e(\rr)}}^2,
\label{eq:OS_strong_conf}
\end{equation}
where $E_\text{P}$ is the Kane energy and $F_{e}(\rr)$ and $F_{h}(\rr)$ are the slowly varying envelopes satisfying the single-particle effective-mass Schr\"{o}dinger equation for the electron and the hole, respectively. The oscillator strength of small quantum dots has therefore an upper limit of $f_\mathrm{max}=E_\text{P}/\hbar\omega$ amounting to 17.4 for GaAs quantum dots at a wavelength of \SI{750}{\nano\meter}.

\subsection{The weak-confinement regime}
Excitons in large quantum dots are bound by the mutual electrostatic attraction between electrons and holes. In this regime, which applies when the mean electron-hole distance, i.e., the exciton Bohr radius, is smaller than the quantum-dot size, the exciton wave function is non-separable implying that the electron and the hole are spatially entangled. The oscillator strength is proportional to the volume of the quantum dot. This effect is sometimes referred to as the giant oscillator-strength effect for quantum dots (GOSQD) and was first considered by Hanamura \cite{hanamura88}. Excitonic enhancements of the light-matter coupling has been studied theoretically in a range of solid-state systems including impurities in bulk semiconductors~\cite{rashba62} and quantum wells~\cite{takagahara86}. The unique feature of the GOSQD is that it occurs for a discrete quantum state.

Our experiments concern investigations of single GaAs interface-fluctuation quantum dots embedded in Al$_{0.33}$Ga$_{0.67}$As as presented in Fig.\ 1\textbf{a} of the main text. Bound excitonic states are obtained by intentionally created monolayer fluctuations in a quantum well, leading to a weak in-plane quantum confinement. The quantum-well thickness is smaller than the bulk exciton Bohr radius of $a_0 = \SI{11.2}{\nano\meter}$~\cite{cardona05} and leads to strong confinement in the growth direction. Exciton enhancement is achieved only within the plane, where the quantum-dot wave function is extended beyond the exciton Bohr radius. We model this as a cylindrically symmetric quantum dot whose slowly varying envelope is separable into in-plane and out-of-plane components and the out-of-plane component is further separable in independent components for the electron and the hole. Hence, the electron-hole motion is correlated in the plane and uncorrelated perpendicularly and the total excitonic wave function can be written in the effective-mass approximation as
\begin{equation}
\Psi(\RR,\rr,\rr_e,\rr_h) = \Psi_\mathrm{X}(\RR,\rr)\psi_e(z_e)\psi_h(z_h)u_e(\rr_h)u_x(\rr_e),
\end{equation}
where $\Psi_\mathrm{X}(\RR,\rr)$ is the in-plane slowly varying envelope, $\psi_e(z_e)$ ($\psi_e(z_h)$) is the out-of-plane envelope function for the electron (hole), $u_e$ ($u_x$) the electron (heavy-hole) Bloch function at the $\Gamma$ point in reciprocal space, $\RR=(m_e\rr_e+m_h\rr_h)/(m_e+m_h)$, $\rr=\rr_e-\rr_h$ the center-of-mass and relative in-plane coordinates of the exciton, and $m_e$ and $m_h$ are the electron and hole effective masses, respectively. The unit-cell Bloch functions contribute to the Kane energy and do not play an important role in our study, which is why only the slowly varying component $\Psi_\mathrm{X}$ is addressed in the main text. The slowly varying envelopes in the growth direction $\psi_{e,h}$ can be accurately computed because the quantum-dot thickness is known precisely and amounts to $L_z=\SI{4.3}{\nano\meter}$ but they play no role for the GOSQD effect, which is governed by the in-plane excitonic envelope $\Psi_\mathrm{X}(\RR,\rr)$. To see this, we first make some realistic assumptions and consider symmetric in-plane parabolic quantum confinement, in which case the excitonic envelope separates into center-of-mass $\chi_\mathrm{CM}(\RR)$ and a relative-motion $\chi_r(\rr)$ components \cite{Ahn2003PRB,Stobbe2012PRB}
\begin{align}
\label{eq:chi_decomp_1}
\Psi_\mathrm{X}(\RR,\rr) &= \chi_\mathrm{CM}(\RR)\chi_r(\rr)\\
\chi_\mathrm{CM}(\RR) &= \sqrt{\frac{2}{\pi}}\frac{1}{\beta}e^{-\abs{\RR}^2/\beta^2} \\
\chi_r(\rr) &= \sqrt{\frac{2}{\pi}}\frac{1}{a_\mathrm{QW}}e^{-\abs{\rr}/a_\mathrm{QW}},
\label{eq:chi_decomp_3}
\end{align}
where $a_\mathrm{QW}$ is the exciton Bohr radius in the quantum well and $\beta$ the in-plane size of the quantum dot. For a perfect two-dimensional system, the exciton Bohr radius is twice as small $a_\mathrm{QW}=a_0/2\simeq \SI{5.6}{\nano\meter}$ leading to a binding energy four times as large. The structure investigated in this work is, however, not a perfect two-dimensional system because the exciton wave function has a non-zero thickness. As argued in Refs.~\citenum{bastard84,greene84}, the binding energy of an exciton in a \SI{4}{\nano\meter}-thick quantum well is only twice larger than in bulk. We therefore consider a value of the two-dimensional Bohr radius $a_\mathrm{QW} \simeq a_0/\sqrt{2} \approx \SI{8}{\nano\meter}$.  For $\beta > a_\mathrm{QW}$, the mean distance between the electron-hole pair ($\approx 2a_\mathrm{QW}$) is smaller than the quantum dot size ($\approx 2\beta$).

\subsection{Relation between the GOSQD effect and single-photon Dicke superradiance}
\begin{figure}
\centering
\includegraphics[width=\columnwidth]{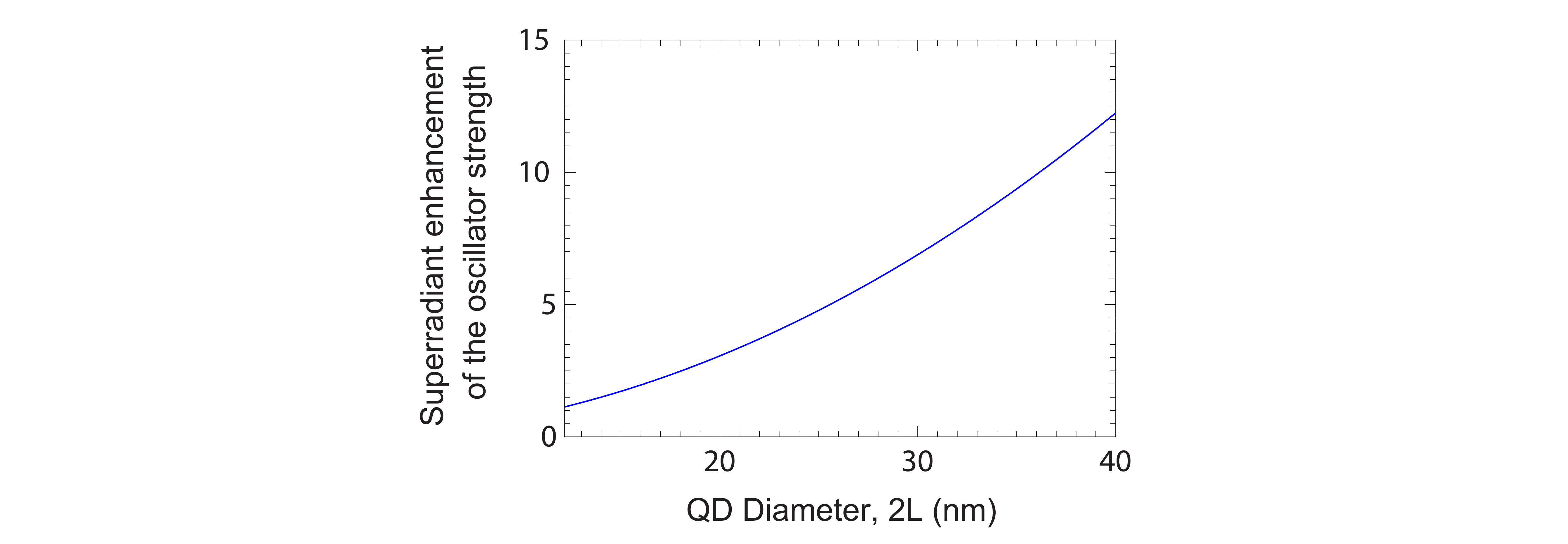}
\caption{Calculated superradiant oscillator strength for an interface-fluctuation quantum dot normalized to the strong-confinement limit of $f_\mathrm{max}=17.4$.
\label{fig:SuperradiantEnhancement_f}}
\end{figure}

The connection to the single-photon Dicke superradiance can be made by noting that, if $\beta > a_\mathrm{QW}$, the center-of-mass motion can be written as a convolution between a function $c_a(\RR)$ capturing the dynamics on the scale of $a_\mathrm{QW}$ and a function $c_s(\RR)$ responsible for the coherent superradiant enhancement, i.e.,
\begin{equation}
\begin{split}
\chi_\mathrm{CM}(\RR) &= c_a(\RR) * c_s(\RR) = \int\mathrm{d}^2\mathbf{P} c_s(\mathbf{P}) c_a(\mathbf{R}-\mathbf{P}) \\
&\approx \sum_j c(\RR_j) c_a(\RR-\RR_j),
\end{split}
\end{equation}
where the last step involves switching the integral to a sum over unit cells and $c$ equals $c_s$ times the discretization area. Consequently, the slowly varying excitonic envelope reads
\begin{equation}
\Psi_\mathrm{X}(\RR,\rr) = \sum_j c(\RR_j)\phi_\mathrm{X}(\RR-\RR_j,\rr),
\label{eq:conv_chi}
\end{equation}
where $\phi_\mathrm{X}(\RR,\rr)=c_a(\RR)\chi_r(\rr)$ is the two-dimensional exciton wave function and is identical to $\phi_\mathrm{X}$ from Eq.~(2) in the main text. The physical meaning of \eqref{eq:conv_chi} is illustrated in Fig.~1\textbf{b} in the main text and the analogy with the Dicke superradiance becomes clear, if compared with Eq.~(1) in the main text. For parabolic in-plane confinement we obtain the following expressions for $c_a$ and $c_s$
\begin{align}
c_a(\RR) &= \frac{1}{\pi a_0^2}e^{-2\abs{\RR}^2/a_\mathrm{QW}^2} \\
c_s(\RR) &= \pi e^{-\abs{\RR}^2/\xi^2},
\end{align}
where $\xi^2=\beta^2-a_\mathrm{QW}^2\approx \beta^2$ for $\beta \gg a_\mathrm{QW}$. Since the phase of $c_s$ is constant throughout the quantum dot, the ground-state exciton is found in a spatial superposition with constructive cooperativity and enhanced coupling to the light field. In the following we quantify the expected superradiant increase in the oscillator strength and, consequently, in the spontaneous-emission rate.

According to Fermi's Golden Rule, the probability of photon emission is proportional to $\abs{\me{0}{\hat{p}_x}{\Psi(\RR,\rr=0,\rr_e,\rr_h)}}^2$. The relative motion is taken to be zero, $\rr=0$, because the exciton can recombine radiatively only if the electron and hole are found at the same spatial position~\cite{Stobbe2012PRB}. After performing the standard procedure of merging the unit-cell Bloch functions into the Bloch matrix element $p_{cv}=V_\mathrm{UC}^{-1}\me{u_x}{\hat{p}_x}{u_e}_\mathrm{UC}$, where the subscript $\mathrm{UC}$ denotes integration over a unit cell, we obtain the following expression for the oscillator strength (compare with \eqref{eq:OS_strong_conf})
\begin{equation}
f = \frac{E_p}{\hbar\omega}\chi_r(0)\abs{\ovI{0}{\chi_\mathrm{CM}(\RR)}}^2 \abs{\ovI{\psi_h(z)}{\psi_e(z)}}^2,
\end{equation}
where the first (second) inner product denotes a two-dimensional (one-dimensional) integration over $\RR$ ($z$). We define the radius of the quantum dot $L=\sqrt{2}\beta$ as argued in Ref.~\citenum{Stobbe2010PRB} and, with the help of Eqs.~(\ref{eq:chi_decomp_1}--\ref{eq:chi_decomp_3}), arrive at the following superradiant enhancement $S$ of the oscillator strength
\begin{equation}
S=\frac{f}{f_\mathrm{max}} = \left(\frac{\sqrt{2}L}{a_\mathrm{QW}}\right)^2 \abs{\ovI{\psi_h}{\psi_e}}^2.
\label{eq:OS_weak_conf}
\end{equation}
Note that this equation is valid only in the weak-confinement regime where $\beta \gg a_\mathrm{QW}$. The electron and hole wave functions in the growth direction can be accurately calculated and we find that $\abs{\ovI{\psi_h}{\psi_e}}^2\approx 0.96$ for the interface-fluctuation quantum dots from the present study. We plot the resulting superradiant enhancement of the oscillator strength in Fig.~\ref{fig:SuperradiantEnhancement_f}. It scales with the quantum-dot area and is a dramatic effect; for realistic quantum dot diameters of \SI{35}{\nano\meter}, the light-matter interaction strength exceeds the upper limit of strongly confined quantum dots with uncorrelated electron-hole pairs by an order of magnitude. %\rsd{The rest of this section sounds like a conclusion. Is this the right place to write all this?} We note that quantum dots are the only obvious solid-state platform where single-photon superradiance can be realized due to their intrinsic zero-dimensional density of states and spatially extended excitons. Importantly, the superradiant state can be created deterministically by a temporally narrow (compared to the decay rate) laser pulse. This is in contrast to non-interacting ensembles of, e.g., atoms where, due to the harmonic spectrum, only the absorption of a single photon can generate single-photon superradiance. Owing to the fermionic nature of the ground-state exciton, quantum dots can be regarded as strong nonlinear systems with unprecedented capabilities of enhancing the light-matter interaction strength at the nanoscale.

\section{II. Sample and experimental setup}
\label{sec:s2}
\begin{figure*}[t!]
	\centering
	\includegraphics[width=.9\columnwidth]{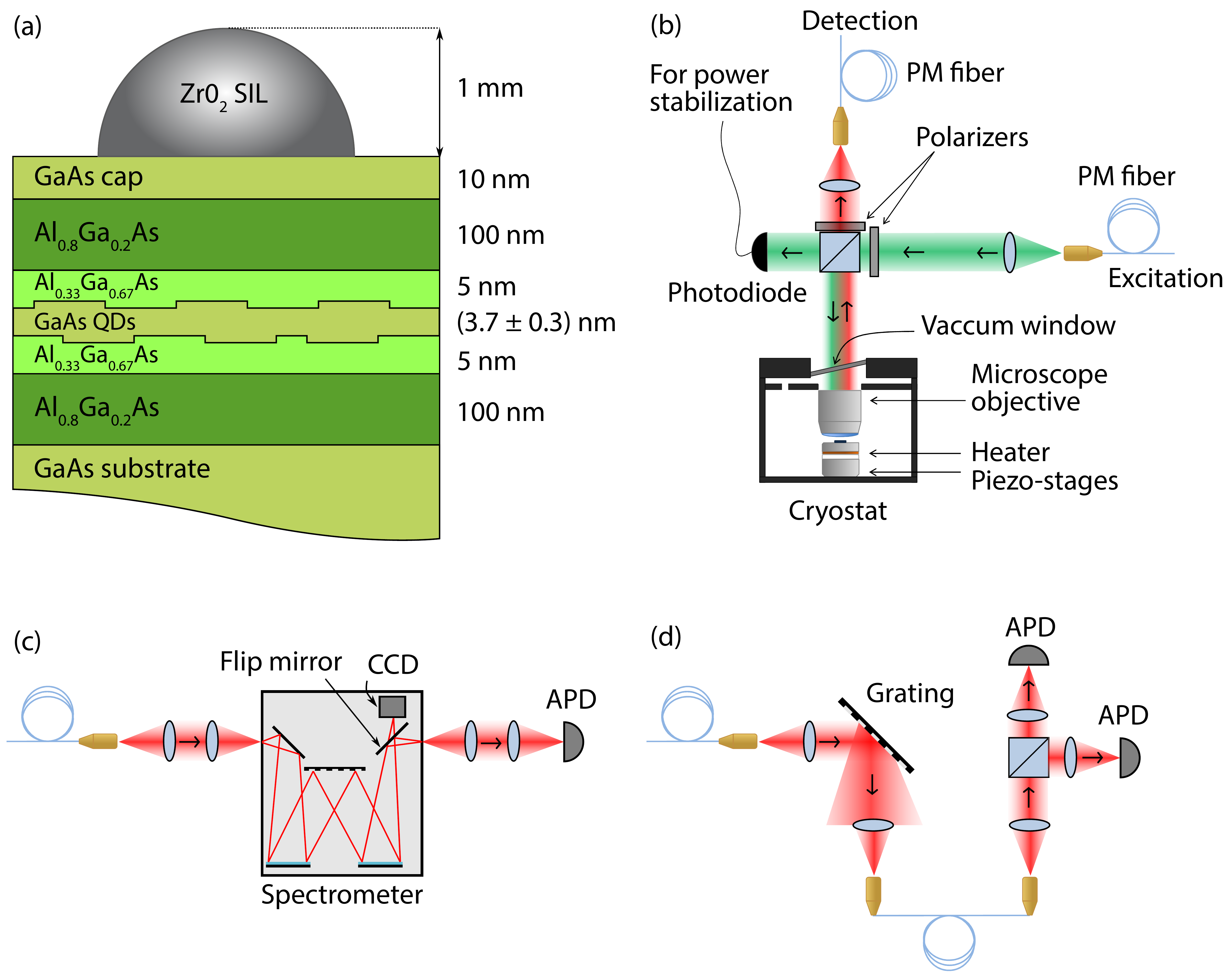}
	\caption{ \label{fig:cutaway_sample}(a) Cutaway profile of the investigated sample (not to scale). Lattice-matched GaAs quantum dots are formed by random fluctuations in the GaAs quantum-well thickness. A zirconia solid-immersion lens enhances the collection efficiency of the setup. (b) Sketch of the optical setup around the cryostat. Optical excitation and collection are performed in a cross-polarized scheme to discriminate between the photoluminescence and specular laser reflection. (c) The emission is sent through a spectrometer before being detected by a CCD for spectral measurements and an APD for time-resolved measurements. (d) For correlation measurements, the quantum dot emission line is spectrally filtered before being directed onto a Hanbury-Brown-Twiss (HBT) setup.}
\end{figure*}

The sample used in our experiment was grown on a GaAs (001) wafer following the procedure developed by Gammon et al.\ \cite{gammon96}. The GaAs interface-fluctuation quantum dots were created by random monolayer fluctuations in the GaAs quantum-well thickness. The GaAs quantum well is surrounded by 5-nm-thick Al$_{0.33}$Ga$_{0.66}$As layers in order to obtain a high-quality interface, and followed by a 100-nm-thick Al$_{0.8}$Ga$_{0.2}$As. The detailed structure is presented in Fig.~\ref{fig:cutaway_sample}(a). A zirconia solid-immersion lens shaped as half a sphere with a radius of \SI{1}{\milli\meter} and refractive index of 2.18 was placed on top of the sample to improve the collection efficiency.

There are several types of optical measurements performed in this study: spectral and time-resolved measurements, and second-order correlation measurements. All of them are carried out in a closed-cycle cryogen-free cryostat as sketched in Fig.~\ref{fig:cutaway_sample}(b). The sample holder is mounted on piezoelectric nanopositioning translation stages. For all experiments, the sample is cooled to a temperature of \SI{7}{\kelvin}. After exiting the single-mode polarization-maintaining (PM) fiber, the excitation beam generated by a picosecond pulsed Ti:Sapph laser is collimated to a diameter of \SI{2}{\milli\meter}. Then, it passes through a thin-film linear polarizer and a 90:10 (transmitted:reflected) beam splitter before being focused on the sample through a microscope objective with a numerical aperture of 0.85. The spatial resolution of the objective was measured to be \SI{1.1}{\micro\meter^2} at a wavelength of \SI{633}{\nano\meter}. The excitation laser is tuned to a wavelength of about \SI{750}{\nano\meter} corresponding either to resonant excitation of continuum states in the quantum well or to $2s$-shell excitation of the quantum dot. The photoluminescence of the investigated ground-state excitons is located around \SI{752}{\nano\meter}. The emission is collimated by the same microscope objective and filtered from the excitation laser by the perpendicularly-oriented thin-film linear polarizer, see Fig.~\ref{fig:cutaway_sample}(b). The beam is then coupled into a PM fiber and guided towards the detection setup.

Spectral measurements are performed by sending the emission to a spectrometer with a groove density and spectral resolution of \SI{1200}{\milli\meter^{-1}} and \SI{25}{\pico\meter}, respectively, and subsequently detected by a charge-coupled device (CCD), see Fig.~\ref{fig:cutaway_sample}(c). After the grating, a mirror can be flipped to direct the emission to an avalanche photo-diode (APD) with a time resolution of \SI{60}{\pico\second} for time-resolved measurements. For correlation measurements, the emission is first filtered by a grating with a groove density of \SI{1200}{\milli\meter^{-1}} before being coupled back into a single-mode PM fiber and directed towards a beam splitter, see Fig.~\ref{fig:cutaway_sample}(d). The grating setup has a spectral resolution of \SI{50}{\pico\meter}. After the beam splitter, two APDs detect coincident counts.

\section{III. Excitation schemes}
\label{sec:s4}
\begin{figure}[t!]
  \centering
  \includegraphics[width=\textwidth]{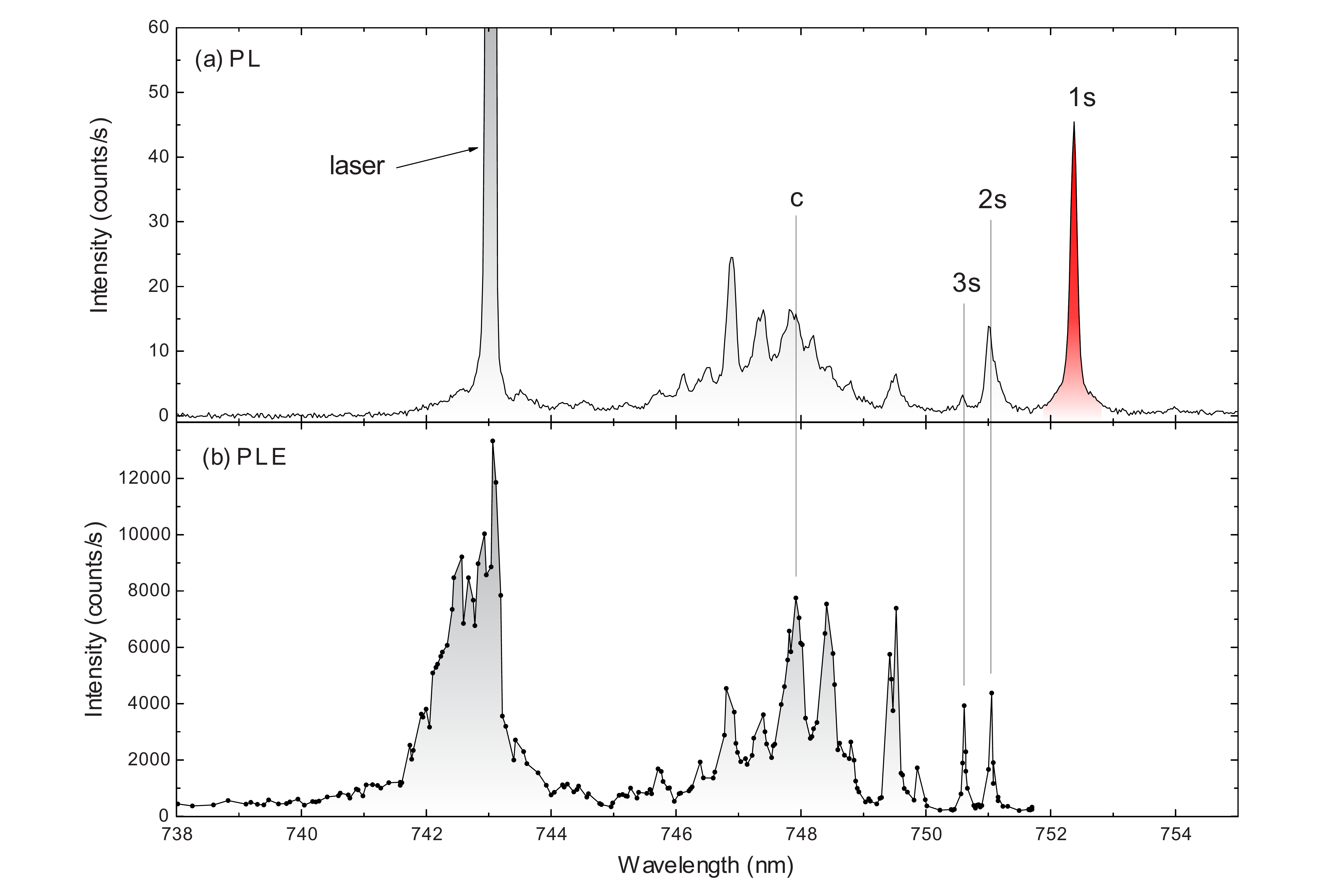}
  \caption{ \label{fig:PLE} Comparison between photoluminescence (PL) and photoluminescence excitation (PLE) spectra from QD2. (a) PL spectrum obtained from exciting the light-hole resonance in the quantum well at 80 nW. The red-shaded area corresponds to the data integrated to obtain the PLE. (b) PLE spectrum performed across the main quantum dot and quantum-well resonances at 400 nW.}
\end{figure}

%The quantum dots are formed by growing a quantum well with a spatially varying thickness~\cite{gammon96,Hours2005PRB}. In the regions where the thickness is larger, bound electronic states are created. Due to weak in-plane confinement, the resulting quantum-dot wave functions are extended in the plane of the quantum well beyond the uncorrelated regime. \rsd{The information up until here is redondant to previous sections, can we remove it?}
In the present work we have studied 9 different interface-fluctuation quantum dots, which are labelled as QD1\ldots QD9. Most of the results presented here stem from QD1 except Fig.~\ref{fig:PLE} (QD2), and Fig.~\ref{fig:QD3} (QD3). All data in the main text stem from QD1 except Fig.\ 2, which was obtained from QD2. We note that all nine quantum dots had very similar properties, cf.\ Table \ref{table:data_all_QDs}.

To acquire a thorough understanding of the level structure of the quantum dot, we measure the electronic density of states in a photoluminescence-excitation (PLE) experiment. The measurement was carried out in continuous-wave mode below the saturation power of the 1s exciton. The laser was scanned stepwise from around 752 nm down to 735 nm where the quantum-dot and quantum-well resonances are present. For each acquired spectrum, the exciton photoluminescence (PL) was fitted by a Lorentzian function and integrated, and the resulting quantity is shown by the height of the quantum-dot line at the given spectral position of the laser as shown in Fig.~\ref{fig:PLE}~(b). We observe a couple of quantum-dot resonances stemming from the 2s- and 3s-shells before the onset of a spectrally continuous absorption of the quantum well. The wavelength used for quasi-continuum excitation is denoted by the C-line in Fig.~\ref{fig:PLE} as well as in the main text. The quantum-well resonances consist of 2D continuous exciton states and quasi-continuum states \cite{guest02}.

The oscillator strength characterizes the coupling strength of a two-level system to light. The quantum dot excitation scheme needs, therefore, to be as clean as possible, such that the environment of the quantum dot is not polluted by phonons and charges that may raise the effective temperature at the quantum dot position and couple the excitonic levels thermally. This requirement is particularly stringent for interface-fluctuation quantum dots owing to the close proximity of the resonances of only a few meV induced by weak quantum confinement. We therefore use the 2s-shell as the main excitation scheme, which also allows us to deterministically prepare the bright superradiant state as explained in the main text. We note that the spectrum remains clean also when pumping through the quantum-well quasi-continuum, cf.\ Fig.\ 1(c-d) in the main text, but the radiative decay rate is found to depend on excitation power, which is attributed to the presence of undesired thermal processes induced by the local phonon bath created by the relaxation of the (many) charge carriers from the quantum well. This excitation scheme is therefore only used for extracting the decay rate of nonradiative processes happening in the quantum dot as discussed below, while the oscillator strength is probed through the 2s excitation.

\section{IV. Measuring the oscillator strength and previous work on GOSQD}
\label{sec:s3}
Experimentally, the oscillator strength can be probed either in absorption \cite{guest02} by extracting the polarizability or in emission by extracting the homogeneous-medium radiative decay rate~\cite{tighineanu13}. The former approach has the advantage that it is unaffected by non-radiative processes but a main drawback is that the quantized character of the excitation is not probed. We therefore study the radiative decay, where the quantized character is encoded in the statistics of the emitted light. The radiative decay is potentially masked by non-radiative and spin-flip processes \cite{Johansen2008PRB}, which are ubiquitous in a solid-state environment. The issue is that a single-exponential decay curve contains contributions from both radiative and non-radiative processes whose sum is measured by fitting the decay curve. Separating radiative from non-radiative processes in such an experiment is therefore not possible. In the present study we address this issue and carefully map radiative and non-radiative contributions using the fine-structure splitting of quantum dots, a technique that results in a bi-exponential decay dynamics of the quantum-dot exciton and is described thoroughly in Refs.~[\citenum{johansen10,tighineanu13,Lodahl2014RMP}].

The decay rate measured in an emission experiment is the total rate,
\begin{align}
\Gamma(\mathbf{r},\omega,{\mathbf{e_p}}) =
\Gamma_{\mathrm{rad}}^{\mathrm{hom}}(\omega)\frac{\rho(\mathbf{r},\omega,{\mathbf{e_p}})}{\rho_{\mathrm{hom}}(\omega)}
+ \Gamma_{\mathrm{nrad}},\label{eq_gamma_tot}
\end{align}
where $\rho(\mathbf{r},\omega,{\mathbf{e_p}})$ and $\rho_{\mathrm{hom}}(\omega)$ are the local density of optical states in the given photonic structure and a homogeneous medium, respectively, $\Gamma_{\mathrm{nrad}}(\omega)$ is the rate of non-radiative recombination, $\hbar \omega$ the emission energy, ${\bf r}$ the position of the quantum dot, and $\Gamma_{\mathrm{rad}}^{\mathrm{hom}}(\omega)$ the radiative decay rate of a quantum dot in a homogenous medium. The oscillator strength is given by \cite{Stobbe2010PRB}
\begin{equation}
	f(\omega) = \frac{6\pi m_0 \epsilon_0 c_0^3}{n(\omega)q^2\omega ^2}\Gamma_{\mathrm{rad}}^{\mathrm{hom}}(\omega),
\label{eq:OS_formula}
\end{equation}
where $n(\omega)$ is the refractive index of the material surrounding the quantum emitter, $\omega$ and $c_0$ the frequency and speed of light, respectively, $\epsilon_0$ the vacuum permittivity, $m_0$ the electron mass, and $q$ the elementary charge. We stress that the oscillator strength is proportional to $\Gamma_{\mathrm{rad}}^{\mathrm{hom}}(\omega)$, which can be very different from the measured decay rate $\Gamma(\mathbf{r},\omega,{\mathbf{e_p}})$. We have precise knowledge of the structure of our sample (see section~\ref{sec:s2}) and, with the help of Ref.~\citenum{paulus00}, calculate a small radiative inhibition of $\rho(\rr,\omega,\mathbf{e_p})/\rho_\mathrm{hom}=0.95$ compared to a homogeneous medium. The fraction of decay events resulting in photon emission is the quantum efficiency defined as
\begin{equation}
	\eta(\omega) = \frac{\Gamma_{\mathrm{rad}}^{\mathrm{hom}}(\omega)}{\Gamma_{\mathrm{rad}}^{\mathrm{hom}}(\omega)+\Gamma_{\mathrm{nrad}}(\omega)}.
\end{equation}

Previous searches for the GOSQD effect was inspired by the prediction by Andreani et al.\ \cite{Andreani1999PRB} that quantum dots in the GOSQD regime may enable reaching the strong-coupling regime of cavity quantum electrodynamics. In some works \cite{Reithmaier2004Nature,Peter2005PRL}, the oscillator strength was estimated from the vacuum Rabi splitting in the strong coupling regime of cavity quantum electrodynamics. Such estimates are inaccurate because multiple quantum dots may couple to the cavity even when they are off resonance due to a (non-Dicke, non-single-photon) collective coupling of multiple quantum dots to the cavity mediated by phonon coupling \cite{Diniz2011PRA,Madsen2013NJP}. In other works, the oscillator strength was estimated from absorption experiments \cite{guest02} but in such experiments the influence of other emitters cannot be ruled out. The oscillator strength has also been estimated from time-resolved measurements \cite{Hours2005PRB,Reitzenstein2009PRL} but, as pointed out above and also noted in Ref.~\citenum{Hours2005PRB}, the non-radiative and radiative processes must be measured independently. It is also crucial to extract properly the radiative decay rate for a homogeneous medium because the local density of optical states is modified significantly in photonic nanostructures even by the presence of nearby planar surfaces. The importance of properly accounting for these effects was highlighted in recent results on large InGaAs quantum dots: in Ref.~\citenum{Reitzenstein2009PRL}, the total decay rate was used to estimate an oscillator strength of $\sim$50 but later measurements showed that non-radiative processes were very significant and that the oscillator strength was $\sim$5 times smaller \cite{Stobbe2010PRB}, i.e., below the GOSQD regime.

\begin{figure}[h!]
  \centering
  \includegraphics[width=0.8\columnwidth]{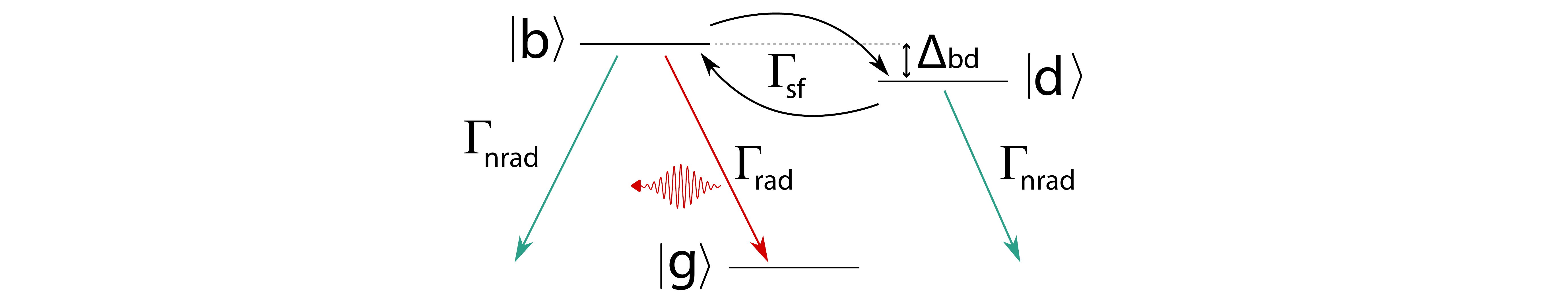}
  \caption{ \label{fig:3level_scheme_X} Three-level exciton scheme with the ground state $\ket{g}$ corresponding to no exciton, and two excited states: bright state $\ket{b}$ and dark state $\ket{d}$ split in energy and coupled through spin-flip interaction with the rate $\Gamma_\mathrm{sf}$. Only $\ket{b}$ can decay radiatively while both states can also decay non-radiatively.}
\end{figure}

In order to extract accurately the oscillator strength of the quantum dots, we use time-resolved spectroscopy on single emitters along with an appropriate model for the bright exciton decay \cite{tighineanu13,Stobbe2010PRB}. Excitons confined in quantum dots arise from the binding of an electron from the conduction band with a heavy-hole from the valence band because compressive strain and confinement shifts the light-hole band to higher energies\cite{Coldren_and_Corzine}. The electron-heavy hole complex has total angular momentum $j=2$ with four possible combinations: $m_j = \{\pm2,\pm1\}$. Excitons with low projected momentum $m_j=\pm 1$ are called bright $\ket{b}$ since they couple to light, while $m_j =\pm 2$ states are dark $\ket{d}$ and  do not couple to light. There is a splitting in energy $\Delta_\text{bd}$ between the two excited states of several hundred $\micro$eV due to electron-hole exchange interaction \cite{Bayer2002PRB}. The level scheme of the exciton is pictured in Fig.~\ref{fig:3level_scheme_X}, where $\ket{g}$ denotes the ground exciton state. The bright exciton can decay both radiatively and non-radiatively while the dark exciton can only decay non-radiatively. The exciton can flip its spin with the rate $\Gamma_\text{sf}$. Spin-flip processes are phonon-mediated, and since $k_BT \gg \Delta_\text{bd}$ for all measurements, the spin-flip rate is assumed to be the same either way between bright and dark excitons. This way, a dark state can contribute to the radiative decay of the bright exciton only if it first undergoes a spin flip to the bright state, thereby providing the slow rate. Finally, the non-radiative rates are taken to be the same for both excitons due to the small energy splitting between the two states \cite{johansen10}.

The population probabilities of bright $\rho_\mathrm{b}$ and dark $\rho_\mathrm{d}$ excitons are governed by the following system of equations\cite{Johansen2008PRB}
\begin{align}
	\frac{\partial\rho_\mathrm{b}}{\partial t} &= -(\Gamma_\mathrm{rad} + \Gamma_\mathrm{nrad} + \Gamma_\mathrm{sf})\rho_\mathrm{b} + \Gamma_\mathrm{sf}\rho_\mathrm{d}, \\
	\frac{\partial\rho_\mathrm{d}}{\partial t} &= -(\Gamma_\mathrm{nrad} + \Gamma_\mathrm{sf})\rho_\mathrm{d} + \Gamma_\mathrm{sf}\rho_\mathrm{b},
\end{align}
and is solved for the bright state $\rho_\mathrm{b}$
\begin{equation}
\rho_\text{b}(t)=\rho_\text{b}(0)\mathrm{e}^{-(\Gamma_\text{rad}+\Gamma_\text{nrad}+\Gamma_\text{sf})t} + \frac{\Gamma_\text{sf}}{\Gamma_\text{rad}}\rho_\text{d}(0) \mathrm{e}^{-(\Gamma_\text{nrad}+\Gamma_\text{sf})t},
\label{eq:bright_X}
\end{equation}
where $ \rho_\text{b}(0) $ and $ \rho_\text{d}(0) $ are the initial populations of dark and bright excitons, respectively. From time-resolved spectroscopy where the decay of the bright exciton is probed, we can retrieve the three decay rates $\Gamma_\text{rad}$, $\Gamma_\text{nrad}$ and $\Gamma_\text{sf}$ through
\begin{align}
\Gamma_\text{rad} &= \Gamma_\text{f} -\Gamma_\text{s} \label{eq:gamma_rad}\\
\Gamma_\text{nrad} &= \Gamma_\text{s} -  \frac{A_\text{s}}{A_\text{f}}\frac{\rho_\text{b}(0)}{\rho_\text{d}(0)}(\Gamma_\text{f} -\Gamma_\text{s}) \label{eq:gamma_nrad}\\
\Gamma_\text{sf} &= \frac{A_\text{s}}{A_\text{f}}\frac{\rho_\text{b}(0)}{\rho_\text{d}(0)}(\Gamma_\text{f} -\Gamma_\text{s}) \label{eq:gamma_sf}
\end{align}

\begin{figure}[t!]
\centering
\includegraphics[width=0.8\textwidth]{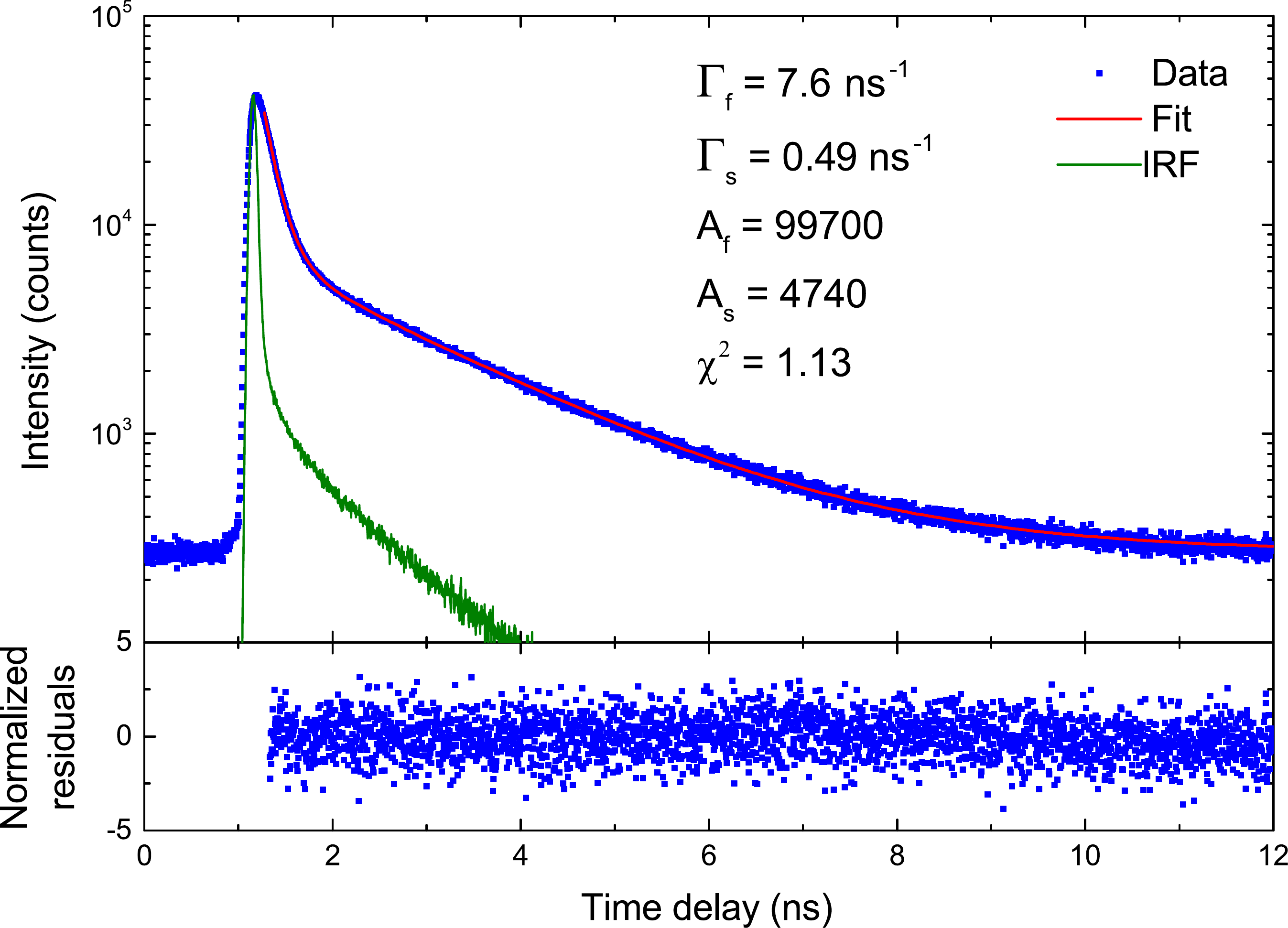}
\caption{Decay curve of QD1 taken under quasi-continuum excitation (blue) along with the bi-exponential fit (red) and the IRF (green). Below are the normalized residuals yielding $\chi^2=1.13$. The extracted parameters for the fit are indicated.}
\label{fig:fitting}
\end{figure}

If the initial preparation probabilities of the bright and dark excitons are known, the radiative rate can be unambiguously extracted from the experimental data~\cite{tighineanu13}. These probabilities are equal when pumping in the quantum well $\rho_\mathrm{b}(0)/\rho_\mathrm{d}(0) \simeq 1$ because carriers with random spins are captured by the quantum dot. If for some reason (e.g., spin-conserving cascade to the ground-state exciton) $\rho_\mathrm{b}(0)/\rho_\mathrm{d}(0) > 1$, we would actually be overestimating $\Gamma_\mathrm{nrad}$ because of \eqref{eq:gamma_nrad}. This means that we are estimating a lower bound to the oscillator strength and quantum efficiency. Quantum-well excitation yields the fast rate $\Gamma_f^\mathrm{hh} = \Gamma_\mathrm{rad}^\mathrm{hh} + \Gamma_\mathrm{nrad} + \Gamma_\mathrm{sf}$, where the three rates correspond to radiative decay, nonradiative decay and bright-to-dark-state spin flip, respectively~\cite{tighineanu13}. We extract $\Gamma_\mathrm{nrad}=\SI{130(3)}{\micro\second^{-1}}$ and $\Gamma_\mathrm{sf}=\SI{360(30)}{\micro\second^{-1}}$, quantities that are not expected to depend on excitation conditions, and this is confirmed experimentally by measuring an excitation-independent slow rate. We note that, due to the presence of thermal processes, $\Gamma_\mathrm{rad}^\mathrm{hh}$ is not related to the oscillator strength but rather to an effective transition strength~\cite{tighineanu13} because more than two levels are involved in the exciton dynamics.

The oscillator strength is measured using 2s-shell excitation, where the fast rate is $\Gamma_f^{2s} = \Gamma_\mathrm{rad} + \Gamma_\mathrm{nrad} + \Gamma_\mathrm{sf}$ and we obtain $\Gamma_\mathrm{rad} = \SI{8.3(1)}{\nano\second^{-1}}$, which is more than four times as fast as the uncorrelated limit of \SI{1.93}{\nano\second^{-1}}, see Fig.~3 in the main text. By taking into account the optical density of states in our structure, this yields a homogeneous-medium decay rate of \SI{8.7}{\nano\second^{-1}}. The experimental proof of SPS is completed in a Hanbury-Brown-Twiss experiment, where the signal is sent to a beam splitter and subsequently detected by two APDs located in the transmission ports of the beam splitter. The resulting coincidence counts on the APDs are shown in Fig.~3 in the main text. We note that 2s-excitation changes the initial preparation condition $\rho_\mathrm{b}(0)\sim 1$ and $\rho_\mathrm{d}(0) \sim 0$ owing to the 2s--1s spin-conserving cascade, which allows deterministic preparation of SPS. This is confirmed in the experimental data, where the quantum-dot decay curves under 2s excitation are reliably fitted with a single-exponential model except in a few quantum dots where the spin-flip rate is fast enough to intermediately create a very small dark-exciton population.

Finally, we discuss the procedure for fitting the decay curves. In time-resolved measurements, the instrument response function (IRF) has to be taken into account when fitting the data. The measured signal $f_\text{meas}(t)$ is the convolution of the input signal $f(t)$ with the IRF $r(t)$ of the APD
\begin{equation}
	f_\text{meas}(t) = \int\limits_{-\infty}^{t}r(\tau)f(t-\tau)\mathrm{d}\tau.
\end{equation}
The laser pulse has a length of 3 ps, and is much shorter than the APD resolution. Therefore, the IRF is measured by collecting light from the laser reflection from the sample surface. Finally, the data are fitted by the aforementioned model (convoluted with the IRF), described by a sum of $N_e$ (either 1 or 2) exponents added to a background value
\begin{equation}
	I_\text{fit}(t) = \text{BG} + \sum_{l=1}^{N_e}b_l\text{e}^{-\Gamma_lt},
	\label{eq:I_fit}
\end{equation}
where $\text{BG}$ is the background level determined by the background counts measured on the APD, and the after-pulsing probability function of the count rate and wavelength. In order to estimate accurately the background level for each measured decay curve, we calibrated the after-pulsing probability close to the quantum dots wavelength for different count rates. The decay curves are fitted by a least-squares approach from which a weighted residual is estimated for each data point $k$,
\begin{equation}
	W_k = \frac{I_\text{meas}(t_k)-I_\text{fit}(t_k)}{\sqrt{I_\text{meas}(t_k)}},
\end{equation}
where $I_\text{meas}(t)$ is the measured data and $I_\text{fit}(t)$ is the fitted value. Finally, the sum of the squared residuals $\chi^2 = (\sum_{k=1}^NW_k^2)/(N-p)$ is minimized to render the best fit to the data, where $N$ is the number of time bins and $p$ is the number of parameters in the model. Figure~\ref{fig:fitting} shows the decay dynamics excited through the quantum-well quasi-continuum and the results of the fitting.

\section{V. Uncertainty of the extracted parameters}
\label{sec:uncertainty}
%Physical quantities can be measured with finite precision. The central message of the present paper is intimately connected to the analysis of time-resolved measurements, which have non-negligible signal-to-noise ratio and finite dynamical range. This naturally leads to uncertainties in the extracted quantities. Another class of uncertainties is related to the sanity of the fitting function, which ideally should reproduce the data well without over-fitting it. The latter may lead to undesired correlations between fitting parameters and large resulting errors. In the following we estimate these uncertainties.

We have carefully addressed the experimental uncertainties for the quantum dot presented in the main text, see also Sec.~\ref{sec:s9}. We have performed the experiment four times in different days and have done a statistical analysis on the extracted quantities, which accounts for experimental errors such as laser-power fluctuations, temperature fluctuations, etc., as well as for the sanity of the fitting routine by randomizing the error signal. The errors of the extracted parameters (decay rates and amplitudes) have been combined in quadrature and propagated using Eqs.~(\ref{eq:gamma_rad}--\ref{eq:gamma_sf}) in textbook fashion~\cite{barlow89}. We obtain an uncertainty of about \SI{2}{\percent} for the oscillator strength and superradiant enhancement as well as for the quantum efficiency.

For the other eight quantum dots a single time-resolved measurement was performed and the uncertainties have been calculated by performing a thorough error analysis on the experimental data. The resulting errors are listed in Sec.~\ref{sec:s9}. In the following we present the method used to extract these uncertainties.

The bi-exponential function from \eqref{eq:I_fit} containing 4 parameters $\vec{p}=(\Gamma_f,\Gamma_s,b_f,b_s)^T$ (note that the background signal BG is a fixed parameter determined by the dark counts and afterpulsing of the APD) is fitted to the raw data with a nonlinear regression employing the Levenberg-Marquardt iterative linearization method~\cite{gregory05}. The regression finds the optimal set of values $\vec{p}_0$ that minimizes the sum of the squared residuals $\chi_\mathrm{min}^2$. To estimate the accuracy of $\vec{p}_0$, the variance-covariance matrix $\textrm{cov}(\vec{p})$ is evaluated at $\vec{p}_0$ using the technique presented in Ref.~\citenum{bard74}. The square root of the diagonal components of this matrix denote the fitting uncertainty of $\vec{p}$ around $\vec{p}_0$, i.e.,
\begin{equation}
\left.\sigma(\vec{p})\right|_{\vec{p}=\vec{p}_0} = \left.\sqrt{\textrm{diag}\left[\textrm{cov}(\vec{p})\right]}\right|_{\vec{p}=\vec{p}_0},
\end{equation}
where $\sigma$ denotes the standard deviation. The entries in $\vec{p}_0$ may have different physical units and it is therefore more convenient to define the dimensionless correlation matrix $\textrm{corr}_{ij}(\vec{p}) = \textrm{cov}_{ij}(\vec{p})/\sigma(p_i)\sigma(p_j)$. Geometrically, the correlation matrix can be envisioned as an ellipsoid in the 4-dimensional parameter space fulfilling $\chi^2 - \chi_\mathrm{min}^2 = 1$. While the diagonal components of the matrix are unity by definition, the off-diagonal components are found between -1 and 1, and contain information about the correlation between different fitting parameters, which quantifies how well the function and the associated parameters are chosen to reproduce the data. A large degree of correlation results in a large redundancy between the parameters, a large fitting error and a small degree of confidence of the fitting procedure. We find that the bi-exponential function models accurately the data judging from the uncertainties being a few percent or less, see Sec.~\ref{sec:s9}, as well as from the low degree of correlation between the parameters that was checked for every fit.

\section{VI. Extracting the size of the exciton wave function}
\label{sec:s6}
\begin{figure}[t!]
	\includegraphics[width=\textwidth]{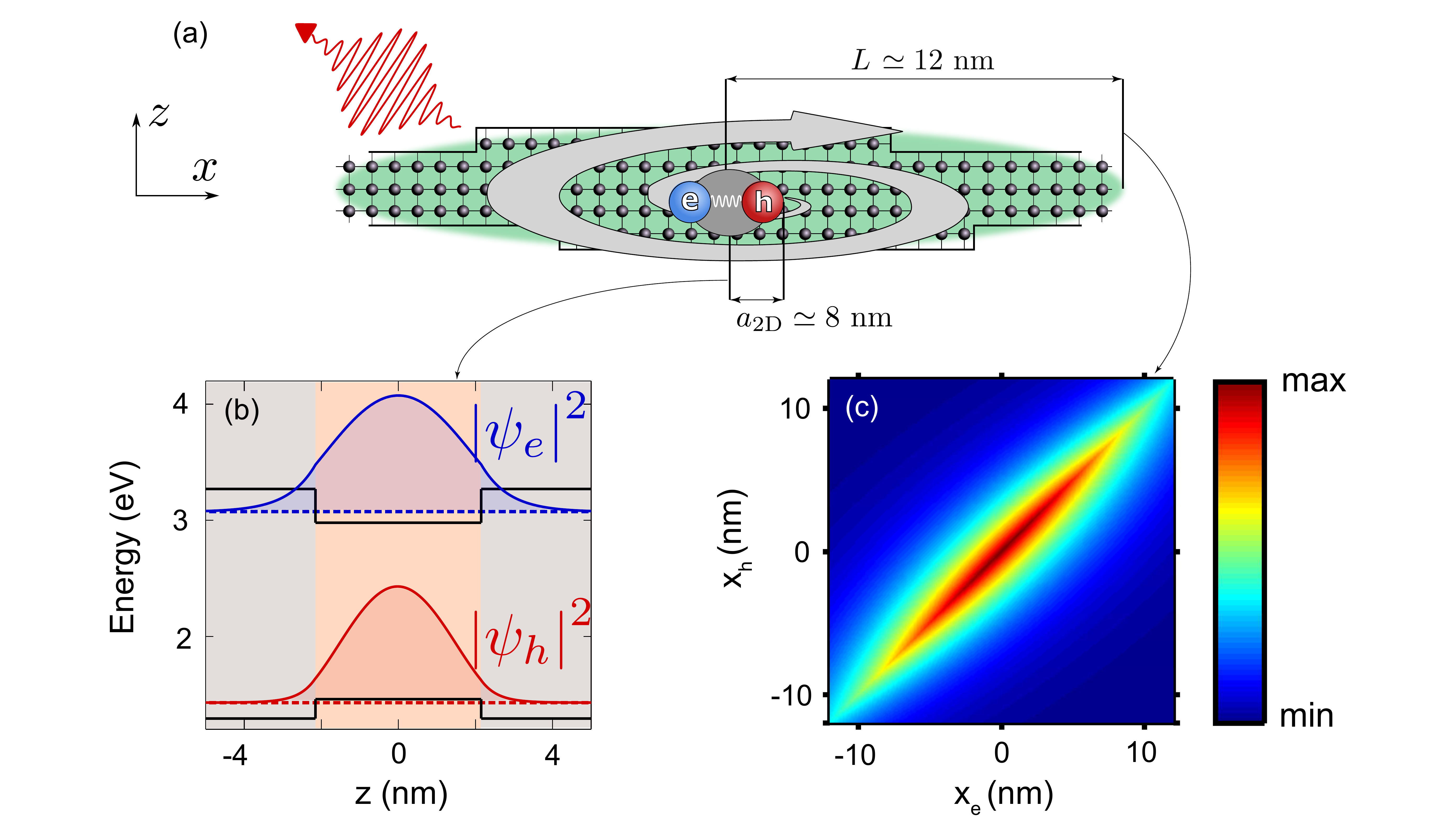}
	\caption{ \label{fig:fig4} (a) Sketch of the interface-fluctuation GaAs quantum dot embedded in an AlGaAs matrix (not to scale). The two-dimensional electron-hole pair (exciton) is coherently spread over the spatial extent of the quantum dot (green area). Exciton enhancement is achieved within the plane (grey arrows), while out of plane cooperative effects are destroyed by the close proximity of the GaAs-AlGaAs potential barrier (black arrows). (b) Band diagram in the transverse direction and the corresponding wave functions of the investigated exciton. The material parameters used in the calculation are taken from Ref.~\citenum{vurgaftman01}. (c) Plot of the in-plane exciton density $\abs{\Psi_\mathrm{X}(x_e,x_h,0,0)}^2$.}
\end{figure}

The presented experimental findings provide insightful information not only about macroscopic properties such as the oscillator strength and quantum efficiency, but also about the nanoscopic structure of the quantum-mechanical wave functions of the quantum dot. The out-of-plane uncorrelated electron and hole wave functions are computed with a tunneling resonance technique~\cite{miller08} and are plotted in Fig.~\ref{fig:fig4}(b) for the investigated quantum dot. While the microscopic structure of the out-of-plane wave functions can be computed because the number of atomic layers in the quantum well is known precisely (see Sec.~\ref{sec:s2}), the in-plane geometry is generally unknown because the quantum-well thickness fluctuations are spatially random. This nanoscopic information is then inferred from the superradiant enhancement of spontaneous emission $S$, where it can be shown (see Sec.~\ref{sec:s1}) that the quantum dot radius $L$ is related to $S$ via
\begin{equation}
L = \frac{a_\mathrm{QW}}{\sqrt{2}}\frac{\sqrt{S}}{\abs{\ovI{\psi_h}{\psi_e}}}.
\label{eq:size_IF_QDs}
\end{equation}
From the measured value $S\simeq 4.3$ an in-plane diameter $2L\simeq \SI{24}{\nano\meter}$ is obtained. The resulting wave function $\Psi_\mathrm{X}(x_e,x_h,0,0)$ is plotted in Fig.~\ref{fig:fig4}(c), where a strong correlation between the electron and hole position within the quantum dot is observed, which gives rise to superradiant emission. Our results emphasize that optical spectroscopy is a robust, non-invasive way of acquiring profound insight into the nanoscopic wave functions of quantum emitters.

\section{VII. The impact of thermal effects on single-photon superradiance}
The central requirement for SPS is quantum dots larger than the exciton radius. The quantization energy $\Delta E_\mathrm{QD}$ scales inversely proportional to the quantum-dot size squared, $\Delta E_\mathrm{QD} \propto L^{-2}$, and, thus, decreases dramatically for large quantum dots. If $\Delta E_\textrm{QD}$ is comparable to the thermal energy $k_\textrm{B}T$, excited states of the exciton manifold become populated with a potentially detrimental effect on the superradiant enhancement of light-matter interaction. This can happen even at cryogenic temperatures owing to the large size of monolayer-fluctuation quantum dots. In the following, we estimate the impact of thermal effects and show that the temperature limits the oscillator strength that can be harvested from a quantum dot. In particular, oscillator strengths larger than about 100 cannot be resolved at the base temperature of our cryostat (\SI{7}{\kelvin}).

\begin{figure}[t!]
	\includegraphics[width=0.8\textwidth]{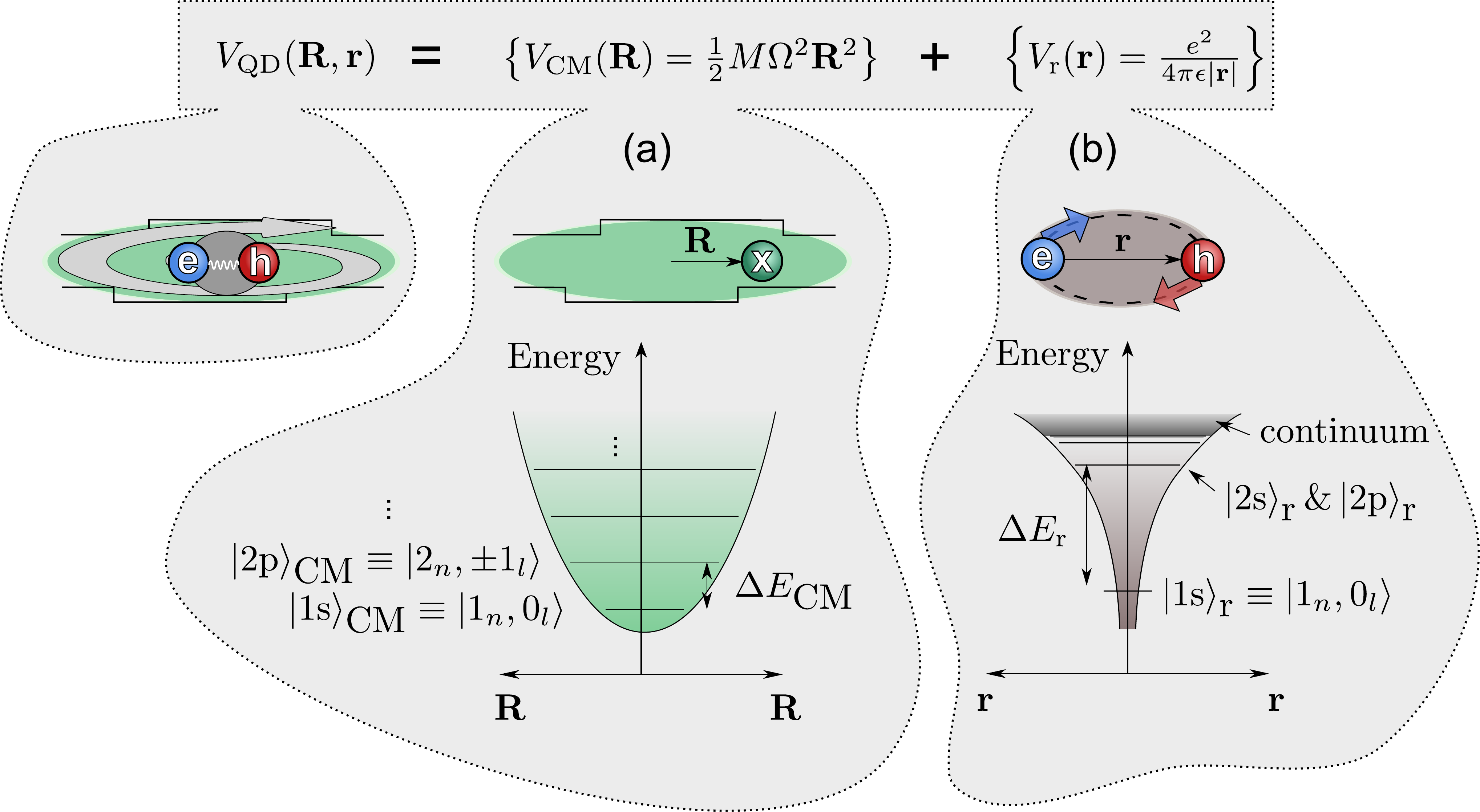}
	\caption{ \label{fig:EnergyLevelsGOS} Decomposition of the in-plane exciton dynamics into (a) a center-of-mass and (b) a relative motion. (a) The center-of-mass dynamics describes the motion of the exciton center of mass in a two-dimensional harmonic potential with equally spaced eigenstates, cf.~\eqref{eq:eigenstates_CM}. (b) The electron-hole electrostatic attraction is captured by the relative-motion dynamics, which is analogous to the two-dimensional Hydrogen with a hyperbolic confinement potential. }
\end{figure}

The process of thermal activation of excited states is intimately connected to the quantum-dot energy structure. We model the monolayer-fluctuation quantum dots as cylindrically symmetric with in-plane parabolic and out-of-plane step-like quantum confinement as explained in Sec.~\ref{sec:s1}. The out-of-plane confinement is strong and thermal effects play no role at cryogenic temperatures. In the following, we address the weak confinement within the plane. Since the exciton dynamics can be decomposed into a center-of-mass (CM) and a relative (r) motion, cf. Eqs.~(\ref{eq:chi_decomp_1}--\ref{eq:chi_decomp_3}), the exciton manifold contains contributions from both. The confinement potential is cylindrically symmetric and the ground state can therefore be denoted as $\ket{1\textrm{s}}_\mathrm{CM}\ket{1\textrm{s}}_\mathrm{r}$ while an excited state can comprise any other combination of the subsets, e.g., $\ket{2\textrm{s}}_\textrm{CM}\ket{1\textrm{s}}_\textrm{r}$, $\ket{2\textrm{s}}_\textrm{CM}\ket{2\textrm{p}}_\textrm{r}$, etc. The relative motion is mathematically equivalent to the two-dimensional Hydrogen problem~\cite{Que1992PRB,sugawara95} and is governed by the mutual electrostatic attraction between the electron and the hole, see Fig.~\ref{fig:EnergyLevelsGOS}, which is why it does not depend on the quantum-dot size. In this subspace, the relevant energy difference $\Delta E_\textrm{r}$ between the ground $\ket{1\mathrm{s}}_\mathrm{r}$ and first excited $\ket{2\textrm{s}}_\textrm{r}$ states equals roughly twice the excitonic Rydberg energy (not four times as for an ideal two-dimensional system, cf. the discussion around Eqs.~(\ref{eq:chi_decomp_1}--\ref{eq:chi_decomp_3})) and amounts to about \SI{8}{\milli\electronvolt}. At cryogenic temperatures, thermal energies are much smaller (below \SI{1}{\milli\electronvolt}) and thermal population of excited states can be safely neglected. We therefore conclude that only the ground state of the relative-motion dynamics $\ket{1\textrm{s}}_\textrm{r}$ is populated, and that thermal effects only affect the spatially extended center-of-mass dynamics.

The center-of-mass motion is described by a particle (exciton) in a two-dimensional harmonic potential $V_\textrm{CM}(\RR)=(1/2)M\Omega^2\RR^2$, cf. Fig.~\ref{fig:EnergyLevelsGOS}(a), where $M$ is the exciton mass and the spring constant $\Omega$ is related to the quantum-dot size $L$ via~\cite{Stobbe2012PRB} $\Omega=4\hbar/ML^2$. The resulting energy eigenstates are given by~\cite{sugawara95}
\begin{equation}
E_{nl} = (2n - \abs{l} - 1)\hbar\Omega,
\label{eq:eigenstates_CM}
\end{equation}
where $n = 1, 2, 3, \ldots$ and $l = 0, \pm 1, \ldots, \pm (n-1)$. The dipole selection rules dictate that states with $l=0$ are bright (superradiant) and all others are dark (subradiant). The present paper studies spontaneous emission from the ground state $\ket{1\textrm{s}}_\textrm{CM} \equiv \ket{1_n0_l}$, which is the only non-degenerate superradiant state of this manifold. For thermal energies $k_\textrm{B} T$ comparable to $\Delta E_\textrm{CM}=\hbar\Omega$, excited states become populated and the relevant figure of merit for light-matter interaction is no longer the oscillator strength $f$, but rather the transition strength $F$ because spontaneous emission no longer involves a two-level system~\cite{tighineanu13}. The transition strength is related to the oscillator strength via a temperature-dependent factor $\mathcal{N}(T)$, i.e., $F(T)=\mathcal{N}(T)\times f$, wherein $\mathcal{N}(T)$ can be regarded as a normalized transition strength and quantifies the distribution of the excitonic population within the center-of-mass subspace. It was shown in Ref.~\citenum{tighineanu13} that for a single excited state, $\ket{e}$, $\mathcal{N}$ is given by
\begin{equation}
\mathcal{N}(T) = \frac{1 + \frac{f_e}{f}\mathcal{B}(T)}{1+\mathcal{B}(T)},
\label{eq:F_2level}
\end{equation}
where $\mathcal{B}(T)=\exp\left( -\Delta E_\textrm{CM}/k_\textrm{B}T \right)$ is the Boltzmann factor and $f_e$ is the oscillator strength of the excited state. If $\ket{e}$ is dark, $f_e\approx 0$, and the temperature is sufficiently high, $\mathcal{B}\approx 1$, the transition strength is half of $f$, i.e., $\mathcal{N} =  1/2$. Here we are interested in finding the threshold temperature at which a certain oscillator strength $f$ of a monolayer-fluctuation quantum dot can still be resolved. This corresponds to the regime in which no excited states are populated and $F \approx f$. Since monolayer-fluctuation QDs have, in general, a large center-of-mass subspace, we generalize \eqref{eq:F_2level} with the help of \eqref{eq:eigenstates_CM} and by noting that the oscillator strength of s-type states equals $f$ while all other states are dark
\begin{figure}[t!]
	\includegraphics[width=0.6\textwidth]{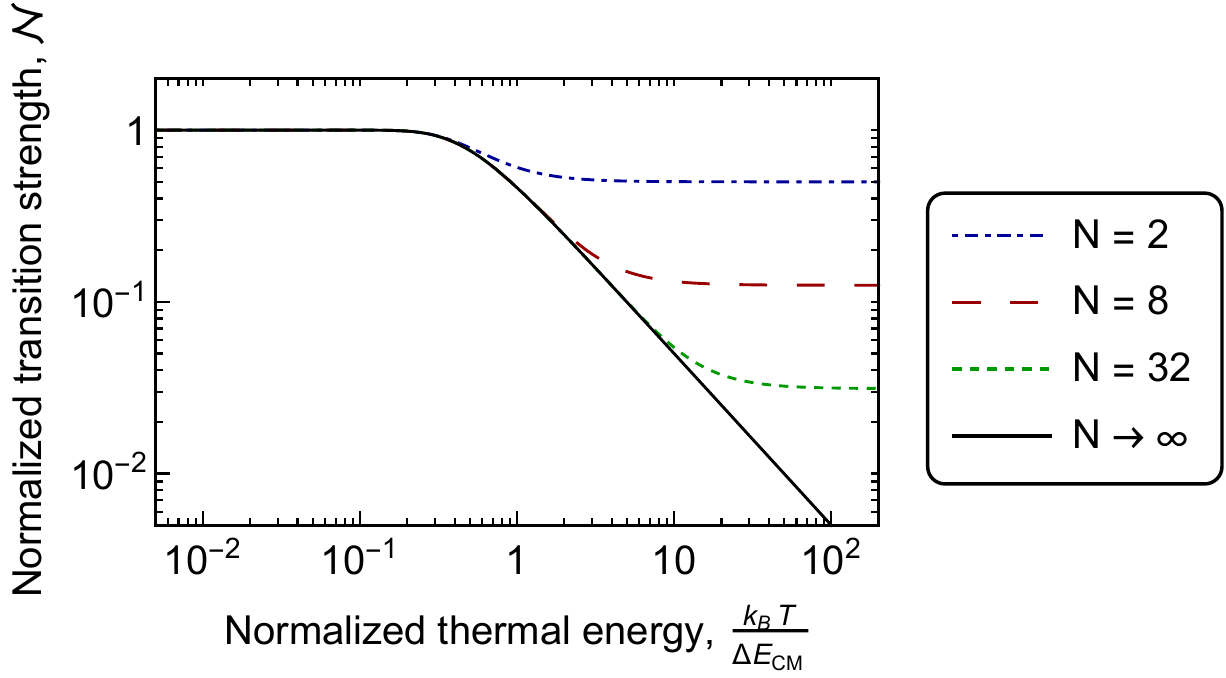}
	\caption{ \label{fig:TransitionStrengthGOS} Plot of the normalized transition strength $\mathcal{N}(T)$ versus normalized thermal energy $k_\textrm{B}T/\hbar\Omega$ for different numbers of bound states $N$ in a monolayer-fluctuation quantum dot. $\mathcal{N}$ starts decreasing monotonically at around $\hbar\Omega \sim 4k_\textrm{B}T$, which decreases light-matter interaction.}
\end{figure}
\begin{equation}
\mathcal{N}(T) = \frac{\textrm{radiative contributions}}{\textrm{all contributions}} = \frac{\sum_{n=1}^N \mathcal{B}^{2(n-1)}}{\sum_{n=1}^N \sum_{l=-(n-1)}^{n-1} \mathcal{B}^{2(n-1)-\abs{l}}},
\end{equation}
where $N$ denotes the number of center-of-mass states. This expression can be evaluated analytically yielding
\begin{equation}
\mathcal{N}(T) = \coth\left( \frac{\hbar\Omega}{2k_\textrm{B}T}N \right) \tanh\left( \frac{\hbar\Omega}{2k_\textrm{B}T} \right),
\end{equation}
and is plotted in Fig.~\ref{fig:TransitionStrengthGOS}. For confinement energies $\hbar\Omega$ smaller than about $4k_\textrm{B}T$, excited states play a negligible role and $F = f$. This is the regime in which the oscillator strength can be reliably measured. Our experiments are in this regime because we observe negligible emission from excited states below saturation. We employ this criterion $\hbar\Omega = 4k_\textrm{B}T$ to estimate the maximum oscillator strength $f_{\textrm{th,max}}^\textrm{sym}$ that can be resolved at a temperature $T$ and obtain (here and in the following we consider the out-of-plane overlap $\abs{\ovI{\psi_h}{\psi_e}}^2 \approx 1$ for simplicity)
\begin{equation}
f_\textrm{max,th}^\textrm{sym} = \frac{4\hbar E_P}{M\omega a_0^2}\frac{1}{k_\mathrm{B}T},
\label{eq:fmaxth}
\end{equation}
which leads to an oscillator strength of 170 at the base temperature of our cryostat of \SI{7}{\kelvin}. This calculation has been performed for an in-plane cylindrically symmetric quantum dot. We evaluate the same expression for a more realistic asymmetric shape with an aspect ratio of 1:$\xi$ with $\xi \geq 1$ and find that the maximum oscillator strength is decreased by $\xi$
\begin{equation}
f_\textrm{max,th} = \frac{f_\textrm{max,th}^\textrm{sym}}{\xi},
\end{equation}
which is equivalent to Eq.~(3) in the main text. We therefore conclude that oscillator strengths larger than about 100 are unlikely to be resolved at the experimental conditions of the present work, and lower temperatures are required for pushing this limit. As demonstrated in Ref.~\citenum{Stobbe2012PRB}, oscillator strengths of 1500 are realistically achievable in monolayer-fluctuation quantum dots with a radius of about \SI{60}{\nano\meter}, where the electric-dipole approximation is reasonably good. Temperatures below \SI{0.8}{\kelvin} would, however, be required to resolve them. For completeness, we note that $\lim_{T\to\infty} \mathcal{N}(T) = 1/N$, i.e., the transition strength is inversely proportional to the number of bound center-of-mass states in the quantum dot at high temperatures.

\begin{figure}[t!]
	\includegraphics[width=0.6\textwidth]{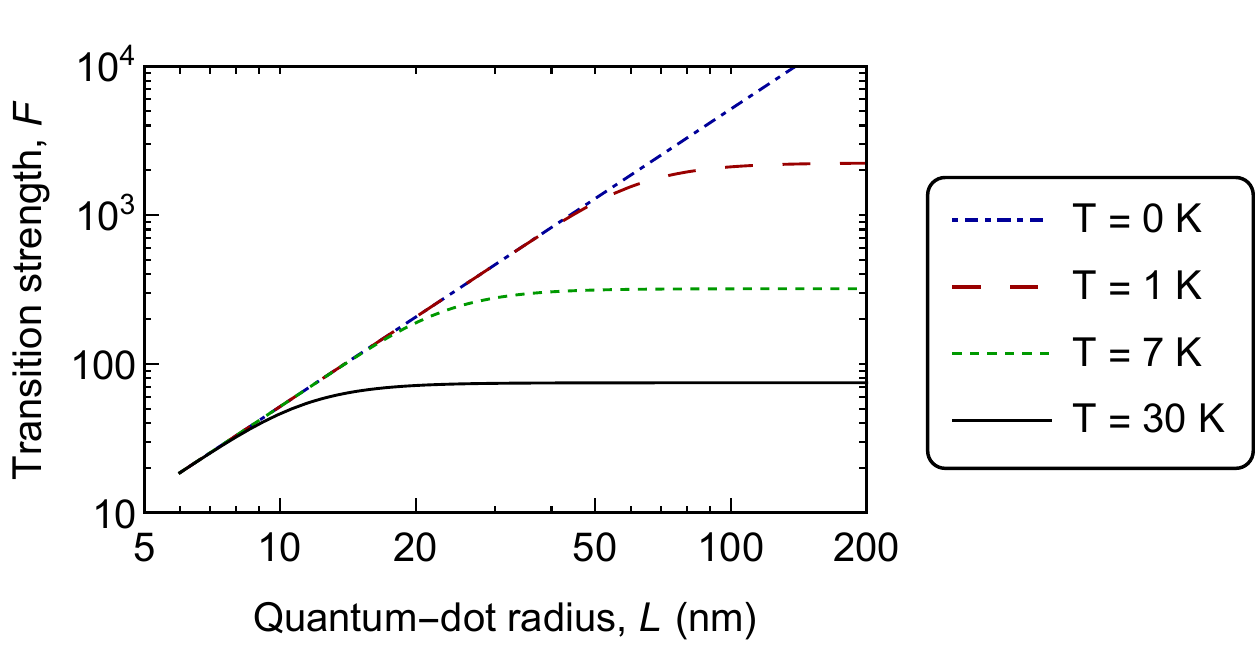}
	\caption{ \label{fig:SuperrEnhGOS} Plot of the transition strength $F$ versus quantum-dot radius $L$ at different temperatures. For small quantum dots, $F$ increases quadratically with $L$ due to superradiance. The transition strength saturates when the thermal energy is larger than the quantum-confinement energy, and becomes independent of quantum-dot size.}
\end{figure}

A natural question arises regarding the transition strength as a function of quantum-dot size $L$. There are two competing processes that scale in opposite directions. On the one hand, the coherent enhancement of light-matter interaction (i.e., SPS) increases with $L^2$, cf.~\eqref{eq:OS_weak_conf}. On the other hand, the normalized transition strength $\mathcal{N}$ decreases monotonically with $\Delta E_\mathrm{CM}^{-1} \propto L^2$ as seen from Fig.~\ref{fig:TransitionStrengthGOS}. From these heuristic arguments it follows that the size dependence cancels out and $F$ saturates for a sufficiently large $L$, which is confirmed in the following quantitative analysis. We assume that the quantum dot has a large center-of-mass subspace $N$, i.e., $\coth\left( \hbar\Omega N/2k_\textrm{B}T \right) \approx 1$, and, from $F = \mathcal{N} f$, obtain
\begin{equation}
F = f_\textrm{max}\left(\frac{\sqrt{2}L}{a_\mathrm{QW}}\right)^2 \tanh\left( \frac{2\hbar^2}{Mk_\mathrm{B}TL^2} \right),
\end{equation}
which is plotted in Fig.~\ref{fig:SuperrEnhGOS}. The transition strength saturates for large $L$ as expected from the aforementioned arguments, and reaches a maximum value of
\begin{equation}
\lim_{L\to\infty}F = \frac{8\hbar E_P}{M\omega a_0^2}\frac{1}{k_\mathrm{B}T},
\end{equation}
which is independent of $L$ and, interestingly, happens to equal $2 f_\textrm{max,th}^\textrm{sym}$. Note that for very large $L \gtrsim \SI{100}{\nano\meter}$, deviations from the electric-dipole approximation, which are not accounted for in this study, further reduce the transition strength~\cite{Stobbe2012PRB}.

\begin{figure*}[t!]
\centering
\includegraphics[width=\textwidth]{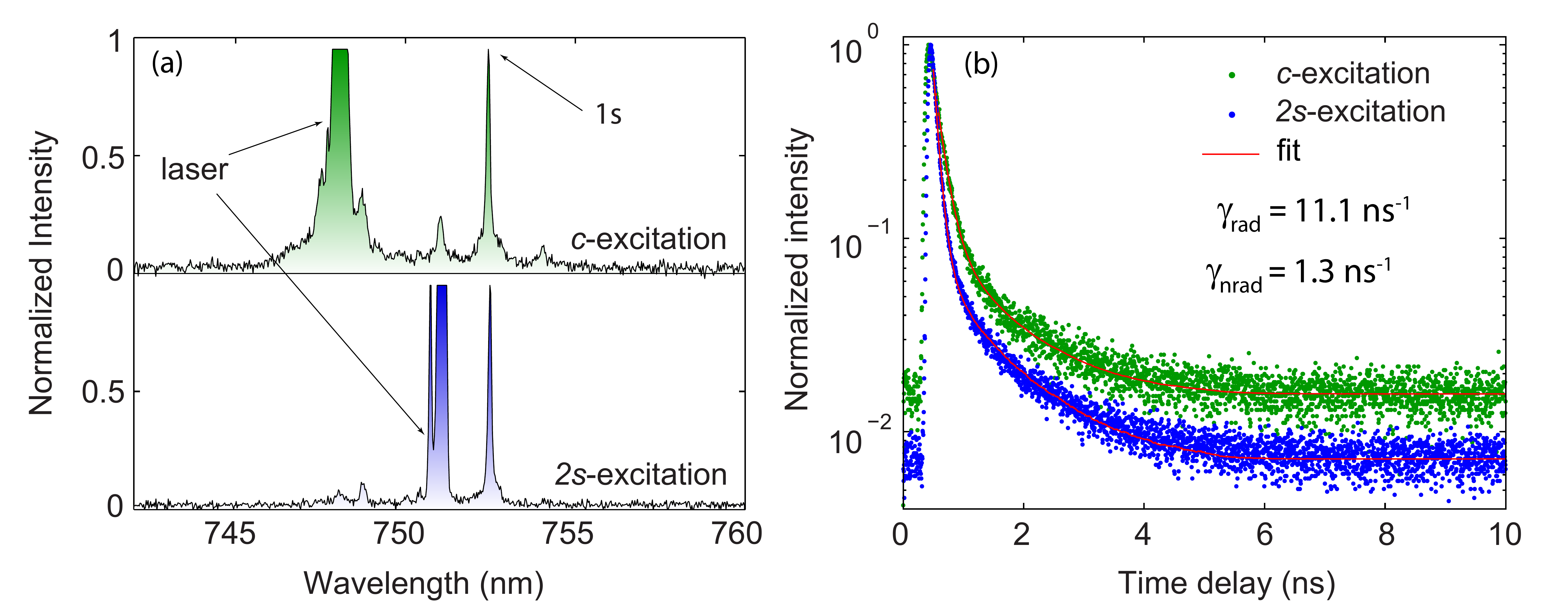}
\caption{Data of QD3 exhibiting the highest measured oscillator strength in this study. (a) Normalized spectra under heavy-hole (green) and 2s-shell excitation (blue) taken at half of the saturation power. (b) Associated decay curves from which we extract an oscillator strength of 96. A good fit to the data obtained with C-excitation (2s-excitation) is obtained with a double (single) exponential function in agreement with the theoretical model.}
\label{fig:QD3}
\end{figure*}

\section{VIII. Results of all measured quantum dots}
\label{sec:s9}

The measurement results for all quantum dots that we studied are presented in Tab.~\ref{table:data_all_QDs}. They represent all measured quantum dots (9 in total), i.e., no data have been discarded after analysis. Figure~\ref{fig:QD3} shows the quantum dot (QD3) exhibiting the fastest decay rate of \SI{11.1}{\nano\second^{-1}}, which corresponds to an oscillator strength of 96. All studied quantum dots have a large oscillator strength with an average value of 76, which constitutes an average superradiant enhancement of 4.4 over the strong confinement limit. The average quantum efficiency is \SI{95}{\percent}. The oscillator strength and quantum efficiency have been estimated with an accuracy of about \SI{2}{\percent} as discussed in Sec.~\ref{sec:uncertainty}.

\begin{table}[h!]
\centering
        \begin{tabular*}{0.75\textwidth}{@{\extracolsep{\fill} } c D{?}{\,\pm\,}{3.3} D{?}{\,\pm\,}{3.3} D{?}{\,\pm\,}{3.3} D{?}{\,\pm\,}{3.3} D{?}{\,\pm\,}{3.3} D{?}{\,\pm\,}{3.3}}
        Quantum dot   &  \multicolumn{1}{c}{$\Gamma_\mathrm{rad}$ (ns$^{-1}$)} & \multicolumn{1}{c}{$\Gamma_\mathrm{nrad}$ (ns$^{-1}$)} & \multicolumn{1}{c}{$\Gamma_\mathrm{sf}$ (\si{\micro\second^{-1}})} & \multicolumn{1}{c}{$f$} & \multicolumn{1}{c}{$\eta$ (\%)} \\
\textbf{QD1} & \textbf{8.3} ? \textbf{0.1} & \textbf{0.130} ? \textbf{0.003} & \textbf{360} ? \textbf{30} & \textbf{72.0} ? \textbf{0.8} & \textbf{99} ? \textbf{2}\\
QD2 & 8.35 ? 0.07 & 0.406 ? 0.008 & 33 ? 2 & 72.0 ? 0.6 & 96 ? 2\\
QD3 & 11.1 ? 0.2 & 1.3 ? 0.2 & 160 ? 30 & 96 ? 2 & 90 ? 2\\
QD4 & 10.5 ? 0.2 & 0.414 ? 0.003 & 47 ? 2 & 90 ? 2 & 96 ? 2\\
QD5 & 7.6 ? 0.2 & 0.93 ? 0.04 & 106 ? 8 & 66 ? 2 & 90 ? 3\\
QD6 & 9.7 ? 0.2 & 0.34 ? 0.02 & 8 ? 2 & 83 ? 1 & 97 ? 2\\
QD7 & 7.13 ? 0.08 & 0.300 ? 0.007 & 13 ? 2 & 61.5 ? 0.7 & 96 ? 2\\
QD8 & 8.13 ? 0.07 & 0.374 ? 0.007 & 15.7 ? 0.9 & 70.1 ? 0.6 & 96 ? 2\\
QD9 & 8.3 ? 0.1 & 0.454 ? 0.007 & 106 ? 5 & 72.0 ? 0.9 & 95 ? 2\\
    \end{tabular*}

    %}
\caption{\label{table:data_all_QDs} Data extracted from time-resolved measurements on all measured quantum dots: radiative decay rate $\Gamma_\mathrm{rad}$, non-radiative decay rate $\Gamma_\mathrm{nrad}$, spin-flip rate $\Gamma_\mathrm{sf}$, oscillator strength $f$, and quantum efficiency $\eta$. QD1: data presented in most of the main article and here, QD2: data of the PLE, QD3: largest oscillator strength. The errors have been evaluated for every quantity as discussed in Sec.~\ref{sec:uncertainty}.} \end{table}

\end{document}